# Effects of Spin-Orbit Resonances and Tidal Heating on the Inner Edge of the Habitable Zone

Christopher M. Colose[1,2], Jacob Haqq-Misra[3], Eric T. Wolf[4,5], Anthony D. Del Genio[6,7], Rory Barnes[8], Michael J. Way[7,9], Reto Ruedy[2]

## Abstract

Much attention has been given to the climate dynamics and habitable boundaries of synchronously rotating planets around low mass stars. However, other rotational states are possible, particularly when higher eccentricity orbits can be maintained in a system, including spin-orbit resonant configurations. Additionally, the oscillating strain as a planet moves from periastron to apoastron results in friction and tidal heating, which can be an important energy source. Here, we simulate the climate of ocean-covered planets near the inner edge of the habitable zone around M to solar stars with ROCKE-3D, and leverage the planetary evolution software package, VPLanet, to calculate tidal heating rates for Earth-sized planets orbiting 2600 K and 3000 K stars. This study is the first to use a 3-D General Circulation Model that implements tidal heating to investigate habitability for multiple resonant states. We find that in the absence of tidal heating, the resonant state has little impact on the inner edge, because for a given stellar flux, higher-order states tend to be warmer than synchronous rotators, but for a given temperature, have drier upper atmospheres. However, when strong tidal heating is present, the rotational component implies a strong dependence of habitable conditions on the system evolution and rotational state. Since tidal and stellar heating both decrease with orbital distance, this results in a compact orbital width separating temperate and uninhabitable climates. We summarize these results and also compare ROCKE-3D to previously published simulations of the inner edge that used a modified version of the NCAR CAM4 model.

[1] Corresponding author c.m.colose@nasa.gov
[2] SciSpace LLC, NASA Goddard Institute for Space Studies, New York, NY 10025, USA
[3] Blue Marble Space Institute of Science, Seattle, WA 98104, USA
[4] NASA NExSS Virtual Planetary Laboratory, Seattle, WA 98195, USA
[5] University of Colorado, Boulder, Laboratory for Atmospheric and Space Physics
[6] Department of Applied Physics and Applied Mathematics, Columbia University, New York, NY 10027
[7] NASA Goddard Institute for Space Studies, New York, NY 10025, USA
[8] Astronomy Department, University of Washington, Seattle, WA 98195-1580, USA
[9] Theoretical Astrophysics, Department of Physics and Astronomy, Uppsala University, Uppsala, SE-75120, Sweden

# 1. INTRODUCTION

The divergent evolution of neighboring solar system planets (Del Genio et al., 2020), and more recently the detection of thousands of extrasolar planets (Udry and Santos, 2007; Borucki et al., 2011), has motivated considerable interest in defining the climatic limits where bodies of liquid water can be thermodynamically stable on the surface of a rocky planet (Kasting et al., 1993). Important to defining the inner edge of the habitable zone (IHZ) is the mixing ratio of water vapor in the stratosphere, which is linked to irreversible planetary water loss.

For low mass stars, the habitable zone lies close to the host star where gravitational interactions despin a planet's rotation (Kasting et al., 1993; see their Figure 16). Differential gravity across the planet's diameter results in the development of a tidal bulge near the sub-stellar and anti-stellar points. Since the material comprising the planet is not perfectly elastic, the bulges associated with the elongated body will lead or lag the substellar point. The stellar gravitational field exerts a torque that transfers angular momentum and dissipates energy, de-spinning the planet if the rotation is faster than the average orbital angular velocity. For a circular orbit, the equilibrium condition is typically synchronization (when the most elongated dimension is always pointed toward the star), in which the rotation and orbital period are equal (1:1 spin-orbit resonance, see e.g., Barnes, 2017; Pierrehumbert and Hammond, 2019 for review). Most 3-D General Circulation Model (GCM) simulations of tidally influenced planets to date have assumed synchronous rotation (e.g., Joshi et al., 1997; Merlis and Schneider, 2010; Carone et al, 2014, 2015, 2016; Kopparapu et al., 2016, 2017; Checlair et al., 2017, 2019; Fujii et al., 2017; Bin et al., 2018; Haqq-Misra et al., 2018; Del Genio et al. 2019b; Komacek and Abbot, 2019; Yang et al., 2019; Salazar et al., 2020; Sergeev et al., 2020).

Synchronous rotation is not an inevitable outcome of tidal locking, however. For planets that have at least moderate orbital eccentricity, that have asymmetries in shape or internal structure, or that are influenced gravitationally by neighboring planets, higher order spin-orbit resonances are possible. For eccentric orbits, the torque will be greater at periapsis than at other points in the orbit, and the rotation rate may fall into a super-synchronous regime. Notably, common scenarios include resonant states in which the rotation period is an integer or half-integer multiple of the orbital period (Goldreich, 1966; Goldreich and Peale, 1967; Dobrovolskis, 2007; Correia et al., 2008; Ferraz-Mello et al., 2008; Makarov, 2012; Rodríguez, 2012; Correia et al., 2014; Barnes, 2017).

In our solar system, it was believed Mercury was synchronized until radar observations (Pettengill and Dyce, 1965) showed that it has a rotation period of ~59 Earth days and an orbital period of ~88 days, corresponding to the 3:2 spin-orbit resonance (Colombo and Shapiro, 1966; see e.g., Noyelles et al. 2014 for a more recent discussion). Venus also exhibits a rotation period similar to its orbital period, although the spin evolution is also affected by atmospheric thermal tides from solar absorption (Ingersoll and Dobrovolskis, 1978; Dobrovolskis and Ingersoll, 1980; Correia and Laskar, 2001) with a possible contribution from Earth (Gold and Soter, 1969; Caudal, 2010). More generally, atmospheric shortwave absorption or presence of a companion may cause even zero eccentricity planets to fall out of the synchronous regime (Leconte et al., 2015), albeit not necessarily in a resonant state.

Exoplanets exhibit a wide range of eccentricities (Xie et al., 2016), and those around low mass stars are expected to be found in either synchronous or asynchronous states (such as spin-orbit resonances), including those sampled by current astronomical observations (Correia et al., 2008; Ribas et al., 2016; Barnes, 2017). Furthermore, tidally-evolved bodies in the habitable zone can exist around stars hotter than the M or K spectral class, even extending to Sun-like stars under certain initial conditions (Barnes, 2017).

Recently, Kopporapu et al. 2017 (hereafter, K17) used a modified version of the Community Atmosphere Model version 4 (ExoCAM) to sample the water loss and runaway greenhouse limits for synchronously rotating planets near the IHZ for a range of stellar hosts from 2600 K to 4500 K. K17 suggest that for all but the coolest M stars, the inner edge of the habitable zone occurs when the "moist greenhouse" limit of significant water loss is reached, rather than the traditional "runaway greenhouse" limit, which requires even stronger instellation that is in excess of the limit at which a moist atmosphere can radiate thermal energy to space (often called the Komabayashi-Ingersoll limit, or Simpson-Nakajima limit, depending on regional or intellectual heritage, see e.g., Pierrehumbert, 2010; Goldblatt and Watson, 2012).

For a given stellar flux, the dependence of the moist greenhouse on the stellar host arises for two reasons. First, cooler stars emit a higher fraction of energy at redder wavelengths, resulting in enhanced near-IR absorption in water vapor bands in addition to decreased Rayleigh scattering. This results in lower bond albedos for M star planets than G star planets, both in controlled experiments (Shields et al., 2013) and expressed in the bulk statistical behavior of a large ensemble of simulations with many varying parameters (Del Genio et al., 2019a). It also drives dayside upwelling circulations that more efficiently moisten the stratosphere, enhancing water loss (Fujii et al., 2017). Secondly, synchronous planets in the habitable zone of a cooler star will have shorter orbital periods (equivalent to shorter rotation periods) than those around hotter and more massive stars, due to Kepler's 3rd law. K17 self-consistently adjusted the rotation period of their ensemble of planets for a given stellar flux and stellar host. This affects the general circulation because at certain rotation periods dynamic regime transitions occur (e.g., Edson et al., 2011; Carone et al. 2014, 2015, 2016; Noda et al., 2017; Haqq-Misra et al., 2018). These determine whether the atmospheric dynamics is quasi-geostrophic baroclinic wave-dominated, quasi-barotropic Hadley cell-dominated, planetary wave-dominated, or day-night circulation-dominated. Each regime has different consequences for the efficacy of cloud shielding of incident starlight on the dayside, thus affecting the instellation flux at which the IHZ is reached (Yang et al., 2014; Way et al., 2018).

Departures from the synchronous regime have not been strongly considered in climate studies but may affect habitability in several ways. First, higher order resonant states imply faster rotation for a given orbital period, thus affecting the dynamical regime a planet may be in. Spin-orbit resonance states also affect the spatial distribution of incident stellar heating (Dobrovolskis, 2007, 2015). Unlike for synchronous rotation, higher order resonances result in all longitudes receiving at least some stellar heating, but for non-zero eccentricity some longitudes receive more heating than others. For integer resonances such as 2:1, there is still one strongly heated face of the planet, and weak heating elsewhere. For a half-integer resonance such as 3:2, there are two preferentially heated regions instead. For the half-integer cases, two orbits are required for the same side of the planet to face the host star, introducing a biennial cycle in the temporal variability of such bodies.

Another consequence of the habitable zone lying within a zone of significant tidal influence is the potential effect of the tidal heating itself on the climate. This occurs for non-circular orbits when the distance from the host changes and the time-varying tidal forces (inversely proportional to the cube of distance) deform the secondary. For example, because of gravitational interactions with Europa, the varying distance between Jupiter and Io changes the degree to which Io is distorted, frictionally heating the jovian moon at a magnitude of ~2-3 W m$^{-2}$ (Veeder et al., 2004) and resulting in intense volcanism at the surface. By comparison, Earth's global-mean geothermal flux, due to radiogenic and primordial heat rather than tidal processes, is ~0.08 W m$^{-2}$ (Pollack et al., 1993).

Because of the larger reservoir of rotational and orbital energy in a star-planet system than a planet-moon system, tidal heating on exoplanets can potentially reach 1-2 orders of magnitude larger values than that of Io. Barnes et al. (2013) coined the term "tidal Venuses" to designate planets that are uninhabitable because their tidal surface heat flux exceeds the critical flux that results in a runaway greenhouse. This concept has been generalized (Barnes and Heller, 2013; Heller and Armstrong, 2014), with "tidal Venuses" as planets whose tidal heat flux alone is sufficient to drive a runaway greenhouse, and "tidal-insolation Venuses" as planets for which the combination of tidal and stellar heating does the same. These papers also classify several less active regimes such as "super-Ios" or "tidal Earths" that fall within a particular range of heating values, such as those encountered on Earth or near a lower limit implied by a tectonically active body.

Only a few GCM studies have previously considered isolated examples of higher order spin-orbit resonance effects on climate (Wordsworth et al., 2010; Yang et al., 2013, 2020; Turbet et al., 2016; Boutle et al., 2017; Del Genio et al., 2019b). To our knowledge, no previous work has incorporated geothermal heating into a 3-D GCM in the context of evaluating IHZ limits. Haqq-Misra and Kopparapu (2014) did report the impact of a 2 W m$^{-2}$ surface heating in a highly idealized GCM for a synchronous rotation planet, while Haqq-Misra and Heller (2018) conducted idealized GCM simulations of tidally-heated exomoons in synchronous rotation with the host planet. Yang et al. (2013) showed that at 2:1 and 6:1 resonance with a static ocean, Bond albedo is lower than it is for synchronous rotation and decreases rather than increases with incident stellar flux, thus destabilizing the climate as the planet approaches the IHZ. Turbet et al. (2016) considered a 3:2 resonance state and static ocean for Proxima Centauri b, but assumed zero eccentricity, which is an unlikely scenario. This choice also removes the effect of time-mean longitudinal maxima and minima in stellar heating. Boutle et al. (2017) simulated the same planet in 3:2 resonance and 0.3 eccentricity, but with a thin static ocean surface, which produced a longitudinal double-eyeball pattern of surface liquid water roughly coincident with the maxima in stellar heating. Wang et al. (2014) found zonally symmetric temperatures for 3:2 and 5:2 resonances with a static ocean, but Dobrovolskis (2015) showed that this was the result of an incorrect spatial pattern of instellation. Del Genio et al. (2019b) performed the first dynamic ocean simulation of a planet in a higher order resonance, showing that despite the double maximum in instellation at 3:2 resonance, the resulting climate has a tropical liquid waterbelt spanning the planet because of the ocean thermal inertia and heat transport. Yang et al. (2020) used a dynamic ocean and focused on the outer edge of the habitable zone by considering the effect of sea ice drift on snowball transitions for nine exoplanets, including a sampling of the 3:2 resonance, also with zero eccentricity.

The aim of this paper is to build upon understanding of the IHZ for a range of stellar hosts by considering a larger set of orbital states and the possible impact of tidal heating associated with them. We will do this by using a different GCM than K17 (Resolving Orbital and Climate Keys of Earth and Extraterrestrial Environments with Dynamics, or ROCKE-3D, see methods section). The scope of the paper is as follows: 1) We will compare results for water loss limits for the synchronous planets explored by K17 in the ExoCAM model to two configurations of ROCKE-3D, one with a fully dynamic ocean, and another with an immobile "slab ocean" (identical to that used in ExoCAM) that does not transport heat horizontally but still provides a source of heat capacity and water vapor to the atmosphere 2) We will extend the analysis to 3:2 and 2:1 spin-orbit resonance planets with non-zero eccentricity 3) We calculate tidal heating rates for planets orbiting two of the lowest mass stars considered in the suite of simulations and demonstrate the significance of tidal heating for the IHZ with a 3-D GCM.

## 2. METHODS

All climate simulations presented employ ROCKE-3D (Way et al., 2017), a three-dimensional GCM developed at the NASA Goddard Institute for Space Studies. ROCKE-3D has been used recently for studies of exoplanetary atmospheres, including the study of ancient Venus (Way et al., 2016; Way and Del Genio, 2020), simulations of Proxima Centauri B (Del Genio et al., 2019b), ocean dynamics and resulting ocean habitats (Olson et al., 2020), effects of varying eccentricity (Way and Georgakarakos, 2017) or obliquity (Colose et al., 2019), snowball transitions on tidally locked planets (Checlair et al., 2019), the vertical transport of water vapor in response to different stellar hosts near the IHZ (Fujii et al., 2017), and is a contributing member of the TRAPPIST-1 Habitable Atmosphere Intercomparison project (Fauchez et al., 2020).

For all runs, ROCKE-3D was configured with an atmosphere at 4°x5° latitude-longitude resolution, and 40 vertical layers with a top at 0.1 mb. All simulations herein consider aquaplanets with a 1 bar atmosphere, most of which consist of $N_2$ and $H_2O$, although we perform some 1% $CO_2$ experiments in section 3.3. All planets considered are assumed to have Earth mass and gravity, with zero obliquity. Most runs use a fully dynamic ocean either 158 or 900 m deep (see below) with the same horizontal resolution as the atmosphere.

We perform three distinct sets of simulations, each of which with the goal of identifying the moist greenhouse transition around different stellar hosts. ROCKE-3D is not suited to explore a true runaway greenhouse state due to limitations in the moist thermodynamics, which assume that water vapor is a dilute component of the atmosphere (see Pierrehumbert and Ding, 2016; Ding and Pierrehumbert, 2016 for a discussion) as well as temperature limitations in the radiative transfer. Instead, we focus on the point at which stratospheric water vapor concentration is high enough to result in appreciable water loss over the lifetime of the planet. We draw attention to the so-called "Kasting limit" (Kasting et al., 1993) where stratospheric mixing ratios are $3\times10^{-3}$ (or a specific humidity of ~3 g kg$^{-1}$) where an Earth ocean can be lost over the age of the Earth.

For the simulations presented in section 3.1, we compare the synchronous rotation planets to those explored in the ExoCAM runs of K17 (see their Table 1). These runs are performed with zero eccentricity and a 1 bar pure $N_2$ atmosphere. For these experiments, a fully dynamic 900 m deep ocean was used. Because K17 used a 50 m slab ocean without ocean heat transport (OHT), section 3.1 also presents results from the q-flux configuration of ROCKE-3D (also a 50 m slab ocean with zero OHT). This allows us to compare the effects of other structural differences between the two models while also analyzing the effect of a circulating ocean.

As in K17, we sample stars based on the BT-SETTL model grid of theoretical spectra at effective temperatures of 2600, 3000, 3300, 3700, 4000, and 4500 K, assuming a stellar metallicity of [Fe/H]=0.0 (see supplementary Figure S1). The orbital periods ($P$) in all cases are a function of the stellar host and instellation, and follow:

$$P\ (Earth\ years) = \left[\frac{M}{M_\odot}\right]^{-0.5} \left[\frac{L/L_\odot}{S0X}\right]^{0.75} \quad (1)$$

Where $M/M_\odot$ and $L/L_\odot$ are the stellar mass and luminosity with respect to the solar value, respectively (values taken from Table 1 of K17), and S0X is the top-of-atmosphere incident stellar flux relative to Earth's value, that we take to be $S_{0,earth}$=1360 W m$^{-2}$. $S_0$ is the incident flux on a perpendicular plane to the incoming energy at the semi-major axis of the planet.

The second and largest set of simulations (section 3.2, summarized in our Table 1) samples the 1:1, 3:2, and 2:1 resonance states for planets orbiting the BT-SETTL stars described above, in addition to planets orbiting a star with a solar spectrum ($T_{eff}$=5785 K). As shown in Table 1, the sampling of S0X is somewhat different than in K17, but follows the same principle of gradually moving planets inward until a moist greenhouse state is obtained. The orbital period is given by Equation (1) and the rotation period is adjusted to give the appropriate resonance.

The second set of simulations are performed without geothermal heating and use an eccentricity of 0.2, similar to the value of Mercury. At this eccentricity, planets in the 3:2 resonance are expected to be prevalent at a high frequency (see e.g., Figure 8 in Dobrovolskis, 2007, Figure 2 of Correia and Laskar, 2009; Figure 1 of Barnes, 2017). The 2:1 resonance is also permitted and becomes even more probable at somewhat higher eccentricity values. The 1:1 resonance is unlikely at $e$=0.2, although we perform simulations at this value for comparison to the other resonant states. For this suite of runs, we use a relatively shallow 158 m deep ocean (the bottom of the 5th ocean layer in ROCKE-3D) in order to allow for faster equilibration (versus the 900m deep option) given the large number of runs analyzed.

In the third category of simulations (Table 2) we consider tidal heating in the form of a prescribed geothermal heat flux, $G$. We refer to "tidal" or "geothermal" or "internal" heating interchangeably in the discussion. Heating rates are calculated using VPLanet (Barnes et al., 2020), a software package designed to simulate various aspects of planetary system evolution; here we use the EqTide module that computes equilibrium tide properties. For the calculations presented here, we adopt a tidal quality factor $Q$=100, Love number of degree 2 equal to 0.3, which are typical values for the terrestrial planets and moons in our Solar System (Henning et al., 2009; Barnes et al., 2013), and assume the planets are Earth size and mass. For our purposes, tidal heating is a function of eccentricity, orbital distance, stellar mass, and rotation rate (we ignore obliquity). Due to the strong dependence on orbital distance, only planets around the 2600 K and 3000 K stars were found to have significant tidal heating rates. Following K17, these have masses of 0.0886 and 0.143 $M_\odot$, respectively. For the remaining stars, $G$ is typically much less than 1 W m$^{-2}$. Therefore, for ROCKE-3D simulations with tidal heating, we restrict our focus to the two lowest mass BT-SETTL stars.

In this last set of runs, we set $e$=0.2 for the 3:2 and 2:1 resonance, and $e$=0.05 for the 1:1 resonance. Planets with low but non-zero values of eccentricity can remain stable in the synchronous state, while still experiencing tidal heating. We emphasize that while the direct effect of eccentricity and rotation period have an important influence on climate, a critical component for interpreting the results in section 3.3 is that $G$ increases with eccentricity and S0X (due to decreasing orbital distance), and all else being equal, the importance of rotational tidal energy results in $G$(2:1 resonance) > $G$(3:2) > $G$(1:1). This will lead to much larger differences between climates at different resonances than in the second set of simulations without tidal heating.

The values of $G$ derived from VPLanet are applied to the bottom of the 158 m ocean in ROCKE-3D. We assume tidally-derived heating is spatially uniform and constant throughout an orbit. In addition to an atmosphere with only $N_2$ as the non-condensable component, we also simulate tidally heated worlds with 1% $CO_2$. High outgassing rates on such planets would likely lead to high $CO_2$ levels, although a self-consistent consideration of the geophysics and chemistry of tidally heated bodies is left for future work.

All ROCKE-3D runs use the SOCRATES radiation package (Edwards, 1996; Edwards and Slingo, 1996). Previous work (K17, Bin et al., 2018) has demonstrated the importance of water vapor absorption in IHZ estimates around low mass stars. SOCRATES uses a two-stream approximation with opacities treated using the correlated-$k$ method based on the HITRAN 2012 lists (Rothman et al., 2013), along with the equivalent extinction method to handle gas species with overlapping absorption features (Amundsen et al. 2017). SOCRATES utilizes "spectral files", which contain tables to run the radiation code including radiation bands, $k$-distributions and continuum absorption for various gases, Rayleigh scattering coefficients, and optical properties of water droplets and ice crystals (see Way et al., 2017 for further details). Spectral files are optimized for various star-planet combinations by fitting the number of Gauss points per spectral interval to a transmission error tolerance, based on an assumed star-atmosphere pairing. Most runs use a spectral file designed for modern Earth-like atmospheres but with an increased number of spectral bands to improve performance for other stellar types (21 shortwave and 12 longwave bands), rather than the default GA7.0 configuration of SOCRATES that is optimized only for the present-day Earth-Sun atmosphere-star pairing (6 shortwave and 9 longwave bands). The increased number of spectral bands increases the precision in water vapor absorption in the near-IR, and is better suited for hotter and more humid atmospheres than Earth, and for M-dwarf host stars that emit increased amounts of near-IR radiation (Yang et al. 2016). For the high $CO_2$ runs, we use spectral files with 43 shortwave and 15 longwave bands instead, which have been optimized for Archean Earth-like atmospheres around a variety of stellar types. All spectral files use the MT_CKD 3.0 water vapor continuum (Mlawer et al., 2012).

Most simulations were run to thermal equilibrium in which the net radiative balance of the planet, $N$ (absorbed shortwave minus outgoing longwave energy at the top of the atmosphere) asymptotes toward and oscillates around -$G$ (zero in the absence of tidal heating). ROCKE-3D uses a calendar system in which every exoplanet orbit is divided into 12 "months" in which the angle (stellar longitude) subtended by each month is approximately the same as the corresponding month for the Earth (see Way et al., 2017 for details). Therefore, "July" is longer than "February," but may still be less than an Earth day for the shortest period orbits. The default output for post-processing is at this monthly timescale, so for short period orbits many hundreds or thousands of "months" would be required to average over natural weather variability. For post-processing, we changed the output frequency to be a function of the orbital period, such that files are produced at ~4-5 Earth year intervals (see data availability) with the requirement that an even number of complete orbits are averaged over. All reported results are based on the mean of the last two files (~8-10 Earth years) once equilibrium was reached. In some cases, simulations near or just beyond the moist greenhouse regime crash due to a numerical instability. In these cases, results are reported based on the last averaging period only. In Table 1 and 2, simulations that crashed when $N$ was both declining toward zero with time and was less than 5 W m$^{-2}$ are denoted by $^*$, while those with larger imbalances at the time of crash (and sometimes growing with time) are denoted by $^{**}$. All runs with a geothermal heat flux (Table 2) either came to equilibrium or crashed with $N$ much greater than -$G$, suggesting these would likely be approaching a runaway greenhouse.

In all runs, the longitude at Periapsis is fixed to 282.9° (measured from vernal equinox). In ROCKE-3D, vernal equinox occurs at a fixed fraction of the year (79.5/365) and because the June-July-August months are longer at high eccentricity, it would take until early April for that fraction of year to pass, which also places periapsis in February.

# 3. RESULTS & DISCUSSION

### 3.1. ROCKE-3D (OHT & q-flux) comparisons with ExoCAM

A summary of the first set of model runs with 1:1 resonance and $e$=0 is provided in Figure 1, with global average surface air temperature shown on the top panel and stratospheric specific humidity shown on the bottom panel. Stratospheric specific humidity is taken from level 36 in ROCKE-3d (~1 mb), although results are not sensitive to this choice because water vapor mixing ratios are quite uniform near the model top. Solid lines indicate calculations performed with the ROCKE-3D GCM, with solid squares indicating cases with the full 900 m deep dynamic ocean and open circles indicating cases with a 50 m slab non-dynamic ocean. This set of models follows the same configuration that was used by K17 for stellar spectra, values of S0X, and corresponding choices of the planet's rotation rate and orbital distance (see Table 1 of K17). The set of calculations performed with the ExoCAM GCM by K17 are also plotted in Figure 1 as dashed lines. This first set of experiments is intended to show the effect of dynamic ocean transport on the IHZ with ROCKE-3D as well as any systematic differences between ROCKE-3D and ExoCAM.

The condition for a moist greenhouse climate is shown by the horizontal dashed black line in the bottom panel of Figure 1, which follows the "Kasting limit" of ~3 g kg$^{-1}$ for a moist greenhouse used by K17. This threshold is determined by the loss of water that occurs as it reaches the stratosphere and is photolyzed, with the escape of hydrogen into space limited by diffusion. Although ROCKE-3D does not simulate atmospheric escape, the ~3 g kg$^{-1}$ limit is an estimate of the stratospheric water vapor mixing ratio at which a planet would lose a water inventory equal to Earth's present oceans to space over a timescale approximately equal to the age of Earth, assuming diffusion limited escape (Hunten, 1973; Kasting et al., 1993). The model simulations in Figure 1 that are above the moist greenhouse limit remain numerically stable but represent climate states that would eventually lose their oceans due to the photolysis of water and subsequent hydrodynamic escape.

Below the moist greenhouse threshold, all ROCKE-3D cases show higher average surface temperatures when the dynamic ocean is included. This is consistent with previous ROCKE-3D calculations, which demonstrated that the enhanced horizontal energy transport from a dynamic ocean provides increased stability against glaciation and higher global average temperatures (Checlair et al. 2019; Del Genio et al 2019b). However, this trend reverses for most cases above the moist greenhouse threshold, with the dynamic ocean cases colder than the slab ocean cases for the 4500 K to 3300 K host stars. As the model crosses the moist greenhouse threshold, the intensification of moist convection drives the tropopause closer to the model top and increases water transport to the stratosphere—which eventually leads to a numerical instability. The warmest (and highest humidity) cases shown in Figure 1 represent the last stable solution possible with the model configuration and do not necessarily represent a physical climate instability. Nevertheless, the rapid increase in temperature and humidity beyond the moist greenhouse threshold indicates that the model has entered a moist climate regime in which its oceans would be prone to rapid loss. Horizontal energy transport by a dynamic ocean provides a mitigating role against this rapid increase in temperature and stratospheric water content, with the 4500 K to 3300 K dynamic ocean cases remaining stable at higher values of S0X with drier stratospheres. This indicates that the dynamic ocean enhances energy transport from the day to night side of the planet, which reduces the magnitude of vertical moisture transport at the substellar point and thereby slows the accumulation of stratospheric water vapor.

The 3000 K and 2600 K simulations show different behavior from the others, with the dynamic ocean cases warmer than the slab ocean cases both before and after the moist greenhouse threshold. This difference occurs because the 3000 K and 2600 K simulations all fall into an intermediate or rapidly rotating dynamical regime, compared to the slow rotation regime of the 3300 K to 4500 K simulations (Haqq-Misra et al. 2018). The intermediate and rapidly rotating regimes show increased energy transport by atmospheric dynamics and reduced moist convection at the substellar point. The dynamic ocean cases thus remain consistently warmer than the slab ocean cases within these dynamical regimes, with the 3000 K simulations showing nearly identical stratospheric specific humidity between the dynamic and slab ocean cases. The ROCKE-3D 3300 K simulations include stable cases within the moist greenhouse regime, whereas K17 found a direct transition to a numerically unstable runaway greenhouse for identical 3300 K simulations using ExoCAM. The set of 2600 K calculations includes four additional cases at higher values of S0X because ROCKE-3D is able to maintain stable temperatures at greater stellar flux values compared to ExoCAM. Stratospheric specific humidity for the 2600 K cases is also lower than the 3300 K case even at the point of instability, which indicates that the 2600 K simulations transition directly to an unstable runaway greenhouse state, bypassing the moist greenhouse as seen in the K17 2600 K cases using ExoCAM.

The comparison of temperature and stratospheric specific humidity with the ExoCAM results from K17 further emphasizes the contributions of a dynamic ocean toward stabilizing a moist climate. Before the moist greenhouse threshold, ExoCAM results are consistently colder and drier than ROCKE-3D results, with the largest differences at the warmer stellar types. Furthermore, the moist greenhouse transition systematically occurs at higher S0X values in ExoCAM than in ROCEK-3D, except for the 2600 K spectral class. Above the moist greenhouse threshold, ExoCAM results are warmer than ROCKE-3D cases for the 4500 K to 3700 K simulations and remain less stable at high S0X values for the 3300 K and 3000 K simulations. The rapid increase in temperature beyond the moist greenhouse limit for ExoCAM shows a steeper slope than most of the ROCKE-3D slab ocean cases, and the ExoCAM stratosphere accumulates moisture more quickly as it approaches the IHZ. This comparison also indicates the presence of a systematic difference in the IHZ between ExoCAM and ROCKE-3D, pronounced most significantly in the coldest 4500 K and 4000 K cases. Such systematic effects likely reflect differences in physical parameterization schemes between the two models, such as the numerical representation of the planetary boundary layer, radiative transfer, moist convection, or cloud formation. Further model intercomparisons will be useful in identifying the specific physical parameterizations that contribute to different model stability limits at the IHZ, and the processes that determine the requisite transport of water vapor into the stratosphere to enter the moist greenhouse.

## 3.2. 1:1, 3:2, 2:1 ROCKE-3D runs at $e$=0.2

Figure 2 shows the distribution of time-mean, top-of-atmosphere incident stellar heating for S0X=1 and $e$=0.2 for each resonant state considered. For the 3:2 resonance, two orbits are required for a symmetric instellation pattern to appear. We confirm in ROCKE-3D that for a 3:2 resonance there are two instellation maxima on opposite sides of the planet that migrate longitudinally between successive orbits, as shown in Dobrovolskis (2015). For the 2:1 resonance there is a single maximum, but heating becomes less concentrated at higher order and all longitudes receive some incident flux.

Supplementary videos 1-3 show the evolution of stellar heating for each resonance for 12 (1:1 and 2:1) or 24 (3:2 resonance) model "months." We note that for 1:1 resonance planets at non-zero eccentricity, the substellar point does not remain fixed as is typically the case for synchronous rotation GCM experiments. Because the rate of the planet's spin about its axis remains fixed while there is variation in the orbital angular velocity, there is oscillation of the zenith angle on the annual cycle. This leads to a libration of the area of maximal heating, in addition to the well-known fluctuation in magnitude of heating between periapsis and apoapsis (see supplementary video 1). In ROCKE-3D, the rocking back and forth of maximal longitudinal heating amounts to an angle of ~23 degrees with respect to the substellar point, consistent with the scaling of ~2 arcsin ($e$) found in Dobrovolskis (2007). For the 2:1 resonance planets, the annual-mean maximum in stellar heating is shifted westward of the 1:1 substellar point due to the timing of periapsis in these runs. Additionally, the movement of the substellar heating is slower near periapsis when the relatively fast orbital velocity "fights" the spin.

When rotation dominates, as on Earth, the sun moves east-to-west. At some moments, such as at periapsis for the 1:1 resonance, it may temporarily migrate west-to-east. Spin-orbit resonant states can lead to a rich palette of intricate sunrise and sunset patterns from the perspective of an observer on the surface (Dobrovolskis, 2007), as occurs on Mercury where at some locations the Sun reverses direction, sets, and then rises again.

Figure 3 shows the global-mean temperature (top row) and 1 mb specific humidity (bottom row) for each resonant state. The expected pattern of increasing temperature for planets around redder stars is maintained at $e$=0.2 and for each resonance. Additionally, long orbital period planets around a solar host (blue line) are the coldest for a given S0X. Figure 4 shows that planetary albedo typically increases monotonically with hotter stars and almost always increases with S0X for each resonance. This is different than the conclusion of Yang et al. (2013) that found a destabilizing albedo response for non-synchronous planets, although we caution that our experimental setup is different from theirs in a number of respects, including the use of a dynamic ocean, non-zero eccentricity, different planetary radius and gravity, and variable rotation period with S0X. They also found a stronger destabilizing response for a 6:1 than a 2:1 resonance, the former we do not simulate.

Before the climate becomes warm enough for a vanishing of sea ice, global ice coverage is highest for the 1:1 resonance and around a solar host (Figure 4, third row), where the bluest spectrum results in the largest surface reflectivity. Higher resonances result in some stellar heating at all longitudes and favor widespread waterbelt states rather than "eyeball" states of liquid water (Del Genio et al., 2019b; Yang et al., 2020). We also simulate a more rapid demise of sea ice as S0X increases for the higher order resonances.

In our simulations, the 3:2 and 2:1 resonance planets remain systematically warmer than the 1:1 counterpart (Figure 2, and see Table 1) for most values of S0X and stellar types. Planetary albedo (Figure 3, top) is lower for the 3:2 and 2:1 resonances, even for warmer climates with little or no sea ice, due to a less developed cloud deck over the sub-stellar point (see Figure S2-S7 for condensed water mass and planetary albedo at all three resonances). Cloud variations for different resonant states and rotation periods are expressed in horizontal variations in reflectivity and outgoing longwave radiation (Figures S8-S10) and may potentially be leveraged by future observations to distinguish between different orbital configurations.

The surface temperature fields for each resonance are shown in Figure 5-7. As discussed in K17, the increasing importance of the Coriolis effect for planets in the habitable zone of smaller stars results in a transition from a circular symmetry about the substellar point (or, in our case, the time-evolving stellar heating for the 2:1 and 3:2 resonances) to a more zonally uniform structure resembling Earth's climate regime. In our simulations, the effect of a dynamic ocean also results in more complex wave-induced structures, particularly for the medium-to-low mass stars. However, for stars of 3700 K and hotter, the imprint of stellar heating on the distribution of temperature is visually apparent in Figures 4-6. For planets around the 3300 K star, the flow begins to smooth temperatures zonally, particularly for the 3:2 and 2:1 resonances where the rotation is faster and the heating is more evenly distributed in longitude. We note that 3300 K synchronous rotation planets are at the edge of the transition points between "slow rotators" and intermediate "Rhines rotators" (Haqq-Misra et al., 2018, see their Figure 4). These have orbital periods of ~19-23 Earth days (Table 1) in our simulations, but the 3:2 and 2:1 planets will rotate faster by a factor 1.5 and 2. For the 2600 and 3000 K planets, the 3:2 and 2:1 planets also feature much more zonal uniformity than the 1:1 planets, and bear closer resemblance to Earth than the "lobster-like" structures observed in other dynamic ocean simulations, such as with Gliese 581g parameters at a rotation rate of 36.7 Earth days (Hu and Yang, 2014) or Proxima b parameters at 11.2 Earth days (Del Genio et al., 2019b). The 3:2 resonance experiments of Del Genio et al. 2019b (7.5 Earth day rotation) also exhibited only modest longitudinal structure in temperature. As discussed in section 3.3, however, planets in these states would likely have significant tidal heating that is not accounted for in the runs from Table 1.

Although the global-mean temperature is hotter for 3:2 and 2:1 planets compared to 1:1 planets when plotted against S0X, the 1:1 simulations have more moist upper atmospheres for a given global-mean surface temperature (see Figure 8, topleft). Therefore, the S0X value required for a moist greenhouse onset has a weak dependence on the spin-orbit resonance (Table 1).

We attribute the enhanced humidity near 1 mb to increased vertical moisture flux that persists at the substellar point. For a given global-mean surface temperature, maximum temperatures tend to be hottest on the 1:1 planets (Figure 8, top right) and the properties of warm, moist air near the surface are communicated to the upper atmosphere. Figure S11 and S12 show longitude-pressure cross sections of meridional-mean temperatures and ω between 30°N and 30°S of selected simulations with similar global-mean surface temperatures. ω is the "omega" vertical velocity in pressure coordinates, or the rate at which a vertically moving air parcel experiences a change in pressure following its motion. As illustrated in Figure S11 and plotted in Figure 8 (bottomleft), 100 mb temperatures are hotter on 1:1 planets when plotted against global-mean temperature and there is enhanced 100 mb vertical moisture flux [Figure 8, bottomright, calculated as -(ω$q$)g$^{-1}$ where $g$ is gravity, $q$ is specific humidity). As discussed in Pierrehumbert and Hammond (2019), horizontal temperature gradients are constrained by the dynamics to be weak in the upper atmosphere away from the frictional surface layer, such that the global properties of air aloft will feel the influence of the convecting area near the substellar point.

These results suggest that for evaluating the stellar flux at which the IHZ is reached, the spectral class of the host star is much more important than the details of the resonant state a planet finds itself in. However, the spin-orbit resonance has a large effect on the transition from an ice-dominated to ice-free state, the distribution of surface temperatures, and is therefore likely to imprint itself on future observables and potentially favorable zones for a biosphere. Future work will be required to assess this at a higher level of detail, including what role continents may have in modifying the pattern of temperatures at higher-order resonances.

In Figure 9, we summarize results obtained for the location of the IHZ for all runs around 4500 - 2600 K stars presented in section (3.1) and (3.2). This figure shows the value of S0X for different stellar types at which a moist or runaway greenhouse is first encountered. These results are shown for the synchronous rotation planets at zero eccentricity from section (3.1) [exoCAM (dashed line), ROCKE-3d slab ocean (solid black line with open circles), ROCKE-3d 900 m dynamic ocean (solid black line with solid squares)] and section (3.2) [ROCKE-3d 158 m dynamic ocean runs at 0.2 eccentricity for 1:1 (orange), 3:2 (purple), or 2:1 resonance (green)]. As discussed in section (3.1), in the synchronous rotation experiments with no eccentricity, the IHZ for ExoCAM coincides with higher stellar fluxes than ROCKE-3D, while the dynamic ocean of ROCKE-3D tends to make the most greenhouse transition more difficult relative to the slab ocean cases.

The difference between the 1:1 resonance cases in section (3.1) and (3.2)(black and orange solid lines) reflect differences in both the eccentricity and ocean depth used. Several sensitivity tests for the 4000 K and 4500 K stars (not shown) suggest the shallow ocean changes the global mean temperature by several degrees when compared to the deeper ocean, but the transition to the moist greenhouse is not strongly affected. Therefore, the eccentricity dominates as the main factor in the IHZ coinciding with lower stellar fluxes. This effect is larger around higher mass stars, likely because the effect on the substellar cloud deck is accentuated. This suggests that the direct effect of eccentricity may not be as important around M-type stars, although we only sampled just one eccentricity value ($e$=0.2) in section 3.2 and future work will be needed to add this component to the parameter space.

Ocean dynamics have previously been found to have a small effect on the climate and phase curves of planets near the IHZ (Yang et al., 2019; that paper sampled 37 and 60 Earth day rotation periods). Figure 9 also indicates that IHZ estimates obtained with the dynamic and slab ocean versions of ROCKE-3D are similar, based on when the upper atmospheric humidity reaches the Kasting limit. The differences between the ROCKE-3D ocean versions are likely within the range of other uncertainties, both related to modeled physical processes as well as the boundary conditions (such as topography) that would exist on real exoplanets. However, we do caution that in some cases we still find large temperature differences near the IHZ between ocean configurations (Figure 1, top). As shown in Figure 9, the influence of ocean dynamics may be comparable to the intermodel spread in IHZ estimates and could be important for threshold cases.

## 3.3. 2600 K and 3000 K stellar host runs with geothermal heating

Figure 10 shows the VPLanet-derived tidal heating rates plotted against the semi-major axis for planets of Earth mass and size orbiting the BT-SETTL 2600 K and 3000 K stars at $e$=0.05 (for the 1:1 resonance) and $e$=0.2 (for 3:2 and 2:1 resonance). The sampled S0X values and corresponding $G$ values are shown in Table 2. We expanded the sampling to lower S0X than in the last section due to the anticipated impact of $G$ on temperatures, and a few of the lowest S0X cases were only performed for high $CO_2$ levels (indicated by bracketed values of temperature or humidity in Table 2). For comparison, in Figure 10 we also show the global-mean incident stellar flux after spherical averaging ($S_0/4$). As shown in Figure 10, the geothermal heating is highest for 2:1 resonance planets (green), followed by 3:2 planets (purple), then the 1:1 planets (orange). The heating rates are also much larger around the 2600 K star than the 3000 K star, and of similar magnitude to the stellar heating for the higher-order resonances. Taken together, the sampled fluxes range from Io-like regimes of ~1 W m$^{-2}$ to >100 W m$^{-2}$.

Figure 11 uses VPLanet to illustrate two example contour maps of tidal heating as a function of rotation period and eccentricity, shown for S0X=0.9 for the 3000 K star, and S0X=0.5 for the 2600 K star. Regions on the contour plots that were sampled in ROCKE-3D have aqua circles. The tidal heating is a complicated function of eccentricity and rotation period, with "ridges" at the 1:1 resonance (red dotted line) and 3:2 resonance. Discontinuities are due to the changes in the signs of phase lags, see Heller et al., 2011 and Barnes et al. 2013 for more details. In general, shorter rotation periods lead to higher values of $G$. As expected, $G$ exhibits a global minimum near the synchronous state and a circular orbit.

Figure 12 summarizes temperature and 1 mb specific humidity information for the suite of runs with geothermal heating. The two panels across each row separate both stellar types, and shows values of (top) temperature or (bottom) 1 mb specific humidity for 1:1 (orange), 3:2 (purple), and 2:1 (green) planets against S0X. 1 bar $N_2$ atmosphere simulations are shown in solid lines connected by squares, and 1% $CO_2$ simulations connected by large circles. Also shown for reference are symbols for the no geothermal runs (all at $e$=0.2) described in the previous section.

For the 1:1 planets around the 3000 K star, the effect of $G$ (~1-2 W m$^{-2}$, see Table 2) is small, and the results bear strong resemblance to the 1:1 simulations with 0.2 eccentricity and zero tidal heating (small red circles). In fact, the 3:2 and 2:1 planets described in the previous section remain warmer than the 1:1 planets with modest tidal heating, although the different eccentricity values may contribute to this. This remains true for S0X=0.8 around the 2600 K star as well, despite a $G$ value of 18.2 W m$^{-2}$. However, for higher S0X the 1:1 geothermal runs around the 2600 K star rapidly become warmer than those in section 3.2, highlighting the strong dependence of $G$ on orbital distance.

For the 3:2 and 2:1 resonant states, the impact of tidal heating is much more pronounced. Temperate climates are achieved at substantially lower stellar fluxes for the 2:1 planets than for 1:1 planets, with 3:2 planets intermediate to these. For example, around the 2600 K star the global-mean temperature for the $N_2$ atmosphere is 13.9 °C at S0X=0.45 for 2:1 resonance, and 1.1 °C at S0X=0.85 for 1:1 resonance. For the 2:1 resonance, the climate approaches the IHZ by S0X=0.5, where the model had a numeric instability at 68 °C and a 1 mb specific humidity that exceeds 3 g kg$^{-1}$ while still warming and far from equilibrium. Because the slope of surface temperature against S0X is much steeper for runs with tidal heating than without, for a 1 bar $N_2$-dominated atmosphere, the orbital width separating a temperate climate from that beyond the IHZ is quite narrow for higher-order resonances. Such planets may transition in and out of moist greenhouse states at various phases of their tidal evolution. In some cases, the climate may bypass an equilibrated moist greenhouse state altogether, and transition from permanently habitable to a runaway. Figure 12 (bottom) shows that for the 2600 K star, upper atmospheric humidity remains low enough for the planet to retain an Earth-sized ocean over several Gyr, but for marginally higher stellar fluxes the model reaches an instability above the Kasting limit with still very large energy imbalances (indicated by small x's).

Although we do not probe the outer edge of the habitable zone in detail, several of our sampled runs highlight that warm climates may be maintained at Martian-like stellar fluxes (around M-dwarf type stars) without a dense atmosphere. For example, our 1% $CO_2$ atmosphere simulations result in global-mean temperatures above freezing for S0X values between 0.45 and 0.5 around the 2600 K stellar host at 3:2 resonance, and 0.4 at 2:1 resonance. Future work will be needed to explore the full width of the habitable zone when allowing for even higher $CO_2$ atmospheres or hydrogen-based atmospheres. However, planets found in eccentric orbits around low mass stars should be expected to have habitable zones at low stellar fluxes.

### 3.4. Additivity of Stellar and Tidal Heating

One of the assumptions implicit in the definitions of "tidal Venuses" and "tidal-insolation Venuses" used by Barnes et al. (2013) and Heller and Armstrong (2014) is that internal heating applied at a planet's surface or the base of the ocean has a climate effect that scales with the magnitude of the heating similar to that which occurs when a planet is subjected to increases in incident stellar radiative flux from above. With ROCKE-3D, we have an opportunity to test the validity of this assumption and to provide clarity to results obtained from simpler climate models.

The stellar constant $S_0$ (defined at the semi-major axis of a planet) is what is prescribed for ROCKE-3D input configuration files. We define a parameter, $S_{0e}$, an equivalent stellar constant such that the absorbed stellar energy of a planet with no geothermal heating would be equal to the time-averaged absorbed energy of a simulation with geothermal heating: That is:

$$A_i = \frac{S_0 f(1-\alpha_G)}{4} + G = \frac{S_{0e} f(1-\alpha_{NG})}{4} \qquad (2)$$

Here, $A_i$ is the annual-mean absorbed energy for a simulation with stellar constant $S_0$ and internal energy flux $G$, and $S_{0e}$ is the variable to be solved for. $f=(1-e^2)^{-0.5}$ is an adjustment factor to account for the increase in annual mean top-of-atmosphere energy input at non-zero eccentricity (~1.02 for $e=0.2$). $\alpha_G$ and $\alpha_{NG}$ is the annual-mean planetary albedo in a simulation with and without geothermal heating, respectively. The former is taken from the last averaging period of the simulations with geothermal heating. We note that the second and third terms in equation (2) are only equal in the annual and global mean, as the prescribed internal heating does not vary with time on the eccentric orbit, and also is horizontally uniform. Because $\alpha_{NG}$ cannot be known *a priori*, we assume $\alpha_{NG} = \alpha_G$ and obtain:

$$S_{0e} = S_0 + \frac{4G}{f(1-\alpha_G)} \qquad (3)$$

It is also possible to equate the energy terms by the total incoming rather than absorbed stellar flux (the denominator in the last term of equation 3 would simply be $f$). However, because the albedo of the atmosphere and surface act to reduce the thermodynamic relevance of a fraction of incident stellar energy, but not $G$, we choose to compare two climates with equivalent $S_{0e}$ by weighting by an approximate albedo contribution to to yield a fairer assessment of the additivity of the two terms. In the discussion below, we discuss results of the geothermal runs from section 3.3 in terms of S0Xe, where S0Xe=$S_{0e}$/1360 W m$^{-2}$. Additionally, we have performed several simulations in which there is no geothermal heating and $S_0$ was chosen to be equal to the $S_{0e}$ of the corresponding geothermal simulation. For these comparisons, the rotation period is kept equal to its geothermal-enabled counterpart.

In Figure 13, we plot results from the geothermal simulations for the 3:2 and 2:1 resonance planets discussed in section 3.3 against S0Xe, as well as the 3:2 and 2:1 simulations without geothermal heating (S0Xe = S0X). The diamond cross data points show the selected runs described above where the stellar flux was increased in order to make up for the missing heating present in its geothermal counterpart. Figure 13 shows that global-mean surface temperature (top) for geothermal runs when plotted against S0Xe tracks very close to the simulations without geothermal heating. This is perhaps expected, since temperature is shackled to the planetary energy balance. However, 1 mb specific humidities for geothermal and no-geothermal runs also bear close resemblance to each other when referenced against identical S0Xe, with the no-geothermal runs remaining moderately more moist for the 2600 K star below the moist greenhouse regime.

For the 3000 K stars where ROCKE-3D reaches a stable equilibrium near the moist greenhouse onset, the IHZ is reached at nearly equivalent values of S0Xe for the planets heated by stellar and tidal heating, or just stellar heating. This result provides assurance that when multiple sources heat a planet, simpler models that consider the total energy input in assessments of the IHZ remain powerful tools for diagnosing the transition to a moist greenhouse.

Figure 14 illustrates the horizontal distribution of temperature in five simulations without geothermal heating (left) but with rotation periods and S0Xe values equal to the geothermal run with the same spectral host and resonance (right). The S0X values used in the left column to obtain identical S0Xe to the runs on the right column are shown alongside the panels. In general, the internal heating at the bottom of the ocean preserves the sense of the large-scale temperature structure, with the largest differences at the highest $G$ values in excess of 100 W m$^{-2}$. For the highest internal heating rate around the 2600 K host ($G$=131.3 W m$^{-2}$) at 2:1 resonance, the ocean develops a "stealth fighter" climate pattern. In all cases, the internal heating acts to reduce the difference between maximum and minimum temperatures. However, these results suggest that the influence of tidal heating on climate may need to be inferred from the orbital configuration of the system rather than any observed climate features.

## 4. Conclusions

We have conducted a suite of simulations with ROCKE-3D for planets at a 1:1, 3:2, and 2:1 spin-orbit resonance around stellar hosts ranging from a 2600 K star to the Sun. We employed VPLanet to derive tidal heating rates for a subset of these planets to assess the importance of internal heat flux for climate. The transition to a moist greenhouse state was found to be only weakly sensitive to the resonant state in the absence of internal heating, despite hotter surface temperature for the 3:2 and 2:1 state for a given stellar flux. We attributed this to a more humid upper atmosphere on 1:1 planets for a given global-mean surface temperature. For simulations with internal heating, however, the importance of rotation to tide-induced heating results in significant differences in climate between resonant states. Because both stellar illumination and tidal heating increase as a planet moves closer to the host star, there is a rapid transition from Earth-like temperate climates to a moist greenhouse with S0X. Tidally influenced planets near the IHZ may be unstable on geologic timescales if they undergo dynamical alterations to eccentricity or rotation period. We also showed that upper atmospheric water mixing ratios are not strongly sensitive to whether a planet is heated by only stellar activity or stellar and tidal activity, given similar total energy inputs.

## Acknowledgments

This work was supported by the National Aeronautics and Space Administration (NASA) Astrobiology Program through collaborations arising from our participation in the Nexus for Exoplanet System Science (NExSS) and the NASA Habitable Worlds Program. Resources supporting this work were provided by the NASA High-End Computing (HEC) Program through the NASA Center for Climate Simulation (NCCS) at Goddard Space Flight Center. M. J. W. acknowledges the support from the GSFC Sellers Exoplanet Environments Collaboration (SEEC), which is funded by the NASA Planetary Science Division's Internal Scientist Funding Model. J.H.M. gratefully acknowledges funding from the NASA Habitable Worlds program under award 80NSSC20K0230 as well as the Virtual Planetary Laboratory Team, a member of the NASA Nexus for Exoplanet System Science, funded under award 80NSSC18K0829. We thank Tom Clune (NASA GSFC) who implemented and provided helpful discussions regarding the ROCKE-3D calendar system.

## Data Availability

# REFERENCES

Amundsen, D.S., Tremblin, P., Manners, J. et al. 2017, *A&A*, 598, A97
Barnes, R. 2017, *CeMDA*, 129, 509
Barnes, R., Mullins, K., Goldblatt, C., et al. 2013, *AsBio*, 13, 225
Barnes, R., Luger, R., Deitrick, R., et al. 2020, *PASP*, 132, 024502
Bin J., Tian F. and Liu L. 2018, *E&PSL*, 492, 121
Borucki, W. J., Koch, D. G., Basri, G., et al. 2011, *ApJ*, 736, 19
Boutle, I. A., Mayne, N. J., Drummond, B., et al. 2017, *A&A*, 601, A120
Carone, L., Keppens, R. and Decin, L., 2014, *MNRAS*, 445, 930
Carone, L., Keppens, R. and Decin, L., 2015, *MNRAS*, 453, 2412
Carone, L., Keppens, R. and Decin, L., 2016, *MNRAS*, 461, 1981
Caudal, G.V., 2010, *JGR*, 115, E07002
Checlair, J., Menou, K., & Abbot, D. S. 2017, *ApJ*, 845, 132
Checlair, J. H., Olson, S. L., Jansen, M. F., & Abbot, D. S. 2019, *ApJL*, 884, L46
Colombo, G., & Shapiro, I. I. 1966, *ApJ*, 145, 296
Colose, C. M., Del Genio, A. D., & Way, M. J. 2019, *ApJ*, 884, 138
Correia, A. C., & Laskar, J. 2009, *Icar*, 201, 1
Correia, A. C. M., Bou´e, G., Laskar, J., & Rodŕıguez, A. 2014, *A&A*, 571, A50
Correia, A. C. M., & Laskar, J. 2001, *Nature*, 411, 767
Correia, A. C. M., Levrard, B., & Laskar, J. 2008, *A&A*, 488, L63
Del Genio, A. D., Kiang, N. Y., Way, M. J., et al. 2019a, *ApJ*, 884, 75
Del Genio, A. D., Way, M. J., Amundsen, D. S., et al. 2019b, *AsBio*, 19, 99
Del Genio, A.D., Brain, D. Noack, L., and Schaefer, L. 2020, Space Science Series, University of Arizona Press, 419

Ding, F., & Pierrehumbert, R. T. 2016, *ApJ*, 822, 24
Dobrovolskis, A. R. 2007, *Icar*, 192, 1
Dobrovolskis, A. R. 2015, *Icar*, 250, 395
Dobrovolskis, A. R., & Ingersoll, A. P. 1980, *Icar*, 41, 1
Edson, A., Lee, S., Bannon, P., et al., 2011, *Icar*, 212, 1
Edwards J. M. 1996 *JAtS* 53 1921
Edwards J. M. and Slingo A. 1996 *QJRMS* 122 689
Fauchez T. J., Turbet M., Wolf E. T. *et al* 2020 *GMD* 13 707
Ferraz-Mello, S., Rodríguez, A., & Hussmann, H. 2008, *CeMDA*, 101, 171
Fujii, Y., Del Genio, A. D., & Amundsen, D. S. 2017, *ApJ*, 848, 100
Gold, T., & Soter, S. 1969, *Icar*, 11, 356
Goldblatt, C. & Watson, A. J. 2012, *RSPTA*, 370, 4197
Goldreich, P., & Peale, S. 1967, *AJ*, 72, 662
Goldreich, P., & Peale, S. J. 1966, *Nature*, 209, 1078
Haqq-Misra, J., & Kopparapu, R. K. 2014, *MNRAS*, 446, 428
Haqq-Misra J., Wolf E. T., Joshi M., et al. 2018 *ApJ* 852 67
Heller, R., & Armstrong, J. 2014, *AsBio*, 14, 50
Heller, R., Leconte, J. & Barnes, R. 2011, A&A, 528, A27
Henning, W.G., O'Connell, R.G., & Sasselov, D.D. 2009, *ApJ*, 707, 1000-1015
Hu, Y., & Yang, J. 2014, *PNAS*, 111, 629
Hunten, D.M. 1973, *J. Atmos. Sci.*, 30, 1481-1494
Ingersoll, A. P., & Dobrovolskis, A. R. 1978, *Nature*, 275, 37

Joshi, M., Haberle, R., & Reynolds, R. 1997, *Icar*, 129, 450
Kasting, J. F., Whitmire, D. P., & Reynolds, R. T. 1993, *Icar*, 101, 108
Komacek, T. D., & Abbot, D. S. 2019, *ApJ*, 871, 245
Kopparapu, R., Wolf, E. T., Arney, G., et al. 2017, *ApJ*, 845, 5
Kopparapu, R., Wolf, E. T., Haqq-Misra, J., et al. 2016, *ApJ*, 819, 84
Leconte, J., Wu, H., Menou, K., & Murray, N. 2015, *Science*, 347, 632
Makarov, V. V. 2012, *ApJ*, 752, 73
Merlis, T. & Schneider, T. 2010, *J. Adv. Model. Earth Syst.*, 2, 13
Mlawer E. J., Payne V. H., Moncet J.-L. *et al* 2012 *RSPTA* 370 2520
Noda, S., Ishiwatari, M., Nakajima, K., et al. 2017, *Icar*, 282, 1
Noyelles, B., Frouard, J., Makarov, V. V., & Efroimsky, M. 2014, *Icar*, 241, 26
Olson, S. L., Jansen, M., & Abbot, D. S. 2020, *ApJ*, 895, 19
Pettengill, G. H., & Dyce, R. B. 1965, *Nature*, 206, 1240
Pierrehumbert, R. T. 2010, *Principles of Planetary Climate* (Cambridge: Cambridge Univ. Press)
Pierrehumbert R. T. and Ding F. 2016 *Proc. R. Soc.* 472 20160107
Pierrehumbert R. T. and Hammond M. 2019 *AnRFM* 51 275
Pollack, H. N., Hurter, S. J., & Johnson, R. R. 1993, *Rev. Geophys.*, 31, 267
Ribas, I., Bolmont, E., Selsis, F., et al. 2016, *A&A*, 596, A111
Rodríguez, A., Callegari, N., Michtchenko, T. A., & Hussmann, H. 2012, *MNRAS*, 427, 2239
Salazar, A. M., Olson, S. L., Komacek, T. D., et al. 2020, *ApJ*, 896, L16
Sergeev, D. E., Lambert, F. H., Mayne, N. J., et al. 2020, *ApJ*, 894, 84
Shields, A. L., Meadows, V. S., Bitz, C. M., et al. 2013, *AsBio*, 13, 715
Turbet, M., Leconte, J., Selsis, F., et al. 2016, *A&A*, 596, A112
Udry, S., & Santos, N. C. 2007, *ARA&A*, 45, 397
Veeder, G. J., Matson, D. L., Johnson, T. V. et al.. 2004, *Icar*, 169, 264
Wang, Y., Tian, F., & Hu, Y. 2014, *ApJ,* 791, L12
Way, M. J., Del Genio, A. D., Kiang, N. Y., et al. 2016, *GRL*, 43, 8376
Way, M. J., Aleinov, I., Amundsen, D. S., et al. 2017, *ApJS*, 231, 12
Way, M. J., Del Genio, A. D., Aleinov, I., et al. 2018, *ApJS*, 239, 24
Way, M. J., & Del Genio, A. D. 2020, *JGR-P*, 125
Wordsworth, R.D., Forget, F., Selsis, F., et al., 2010, *A&A*, 522, A22.
Xie, J.-W., Dong, S., Zhu, Z., et al. 2016, *PNAS*, 113, 11431
Yang J., Abbot D. S., Koll D. D. et al. 2019, *ApJ*, 871 29
Yang, J., Boué, G., Fabrycky, D. C., & Abbot, D. S. 2014, *ApJL*, 787, L2
Yang, J., Cowan, N. B., & Abbot, D. S. 2013, *ApJL*, 771, L45
Yang, J., Ji, W., & Zeng, Y. 2020, *NatAs*, 4, 58
Yang, J., Leconte, J., Wolf, E.T. et al. 2016, *ApJ*, 826, 222

# FIGURES

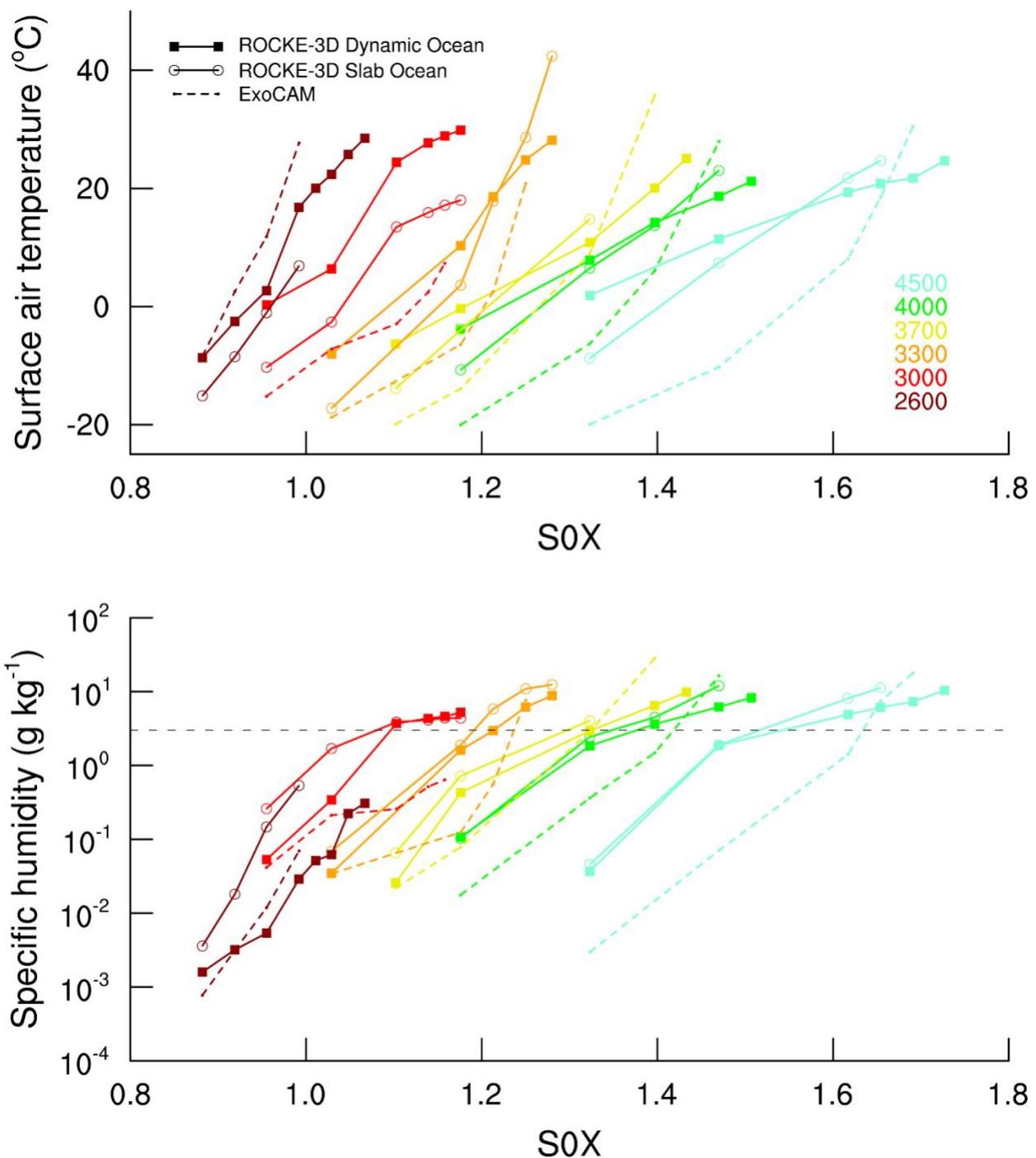

Figure 1. (Top) Global-mean surface air temperature (°C) vs. S0X for 1:1 resonant planets with zero eccentricity and no geothermal heating using (filled squares) ROCKE-3D with dynamic ocean heat transport, (open circles) q-flux version of ROCKE-3D with zero ocean heat transport, and (dashed lines) previously published results using ExoCAM. Results are shown for planets orbiting stars from 2600 K (dark red line) to the 4500 K (cyan line). (Bottom row) 1 mb specific humidity (g kg$^{-1}$) vs. S0X. Horizontal dotted line is plotted at the 3 g kg$^{-1}$ Kasting limit discussed in text.

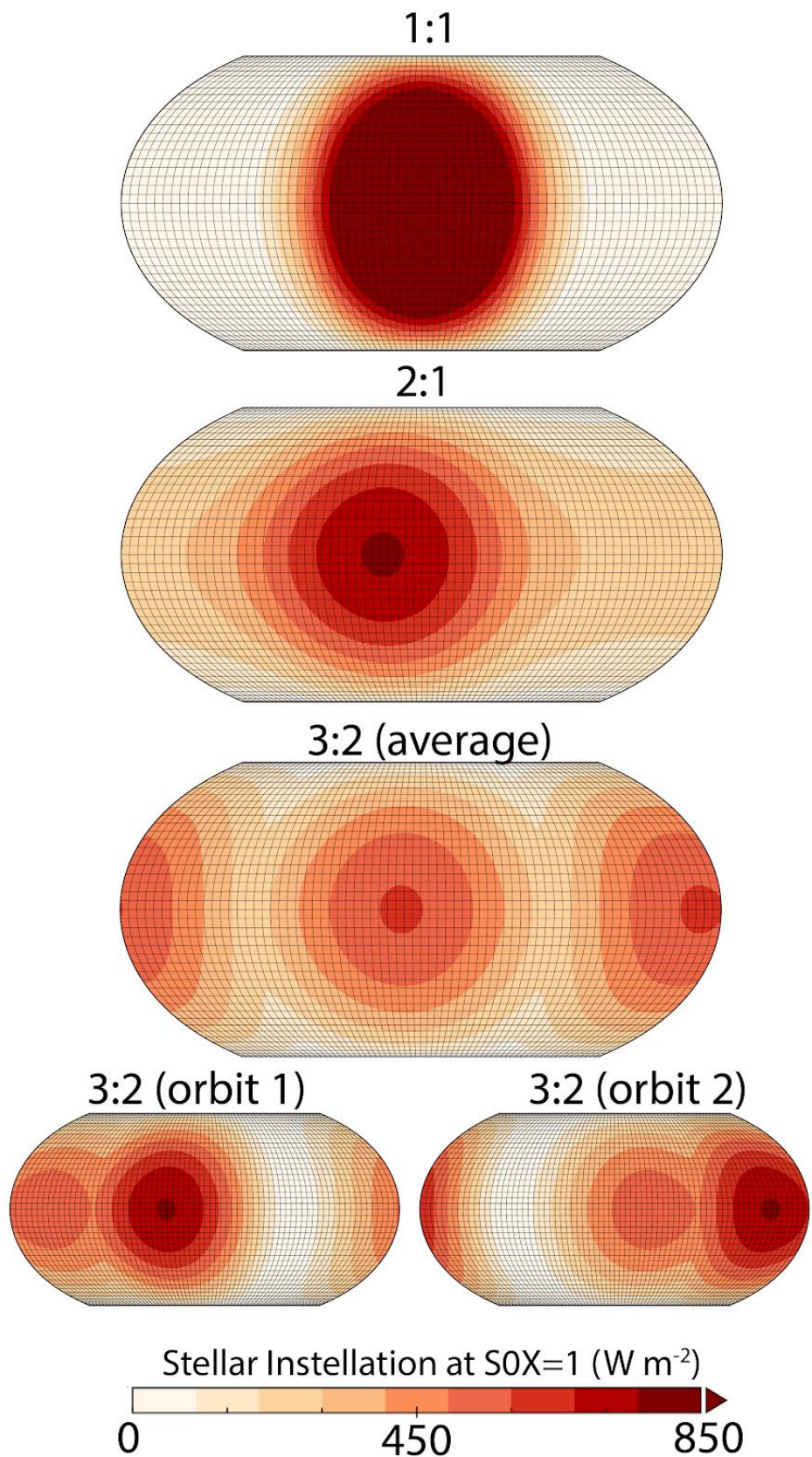

Figure 2. Climatological top-of-Atmosphere instellation pattern for (top row) 1:1, (second row) 2:1, and (third row) 3:2 resonant cases at S0X=1 (1360 W m$^{-2}$) at 0.2 eccentricity. Bottom row shows instellation pattern for every other orbit for the 3:2 resonance.

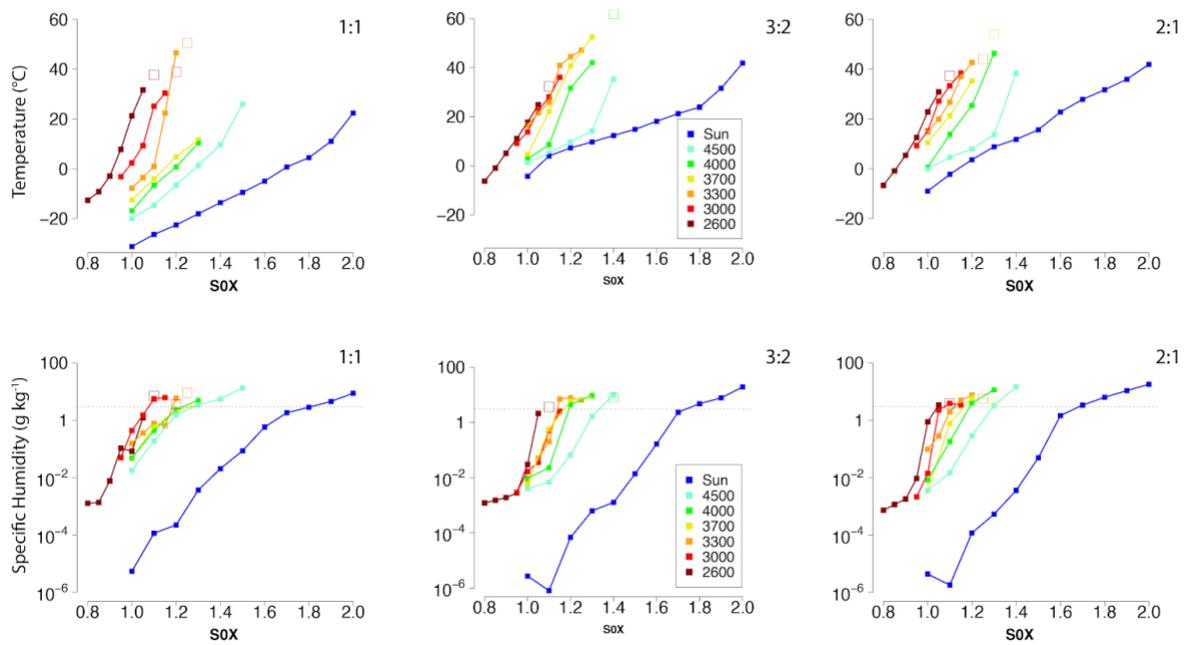

Figure 3. (Top row) Global-mean surface air temperature (°C) vs. S0X for (left) 1:1 (middle) 3:2 (right) 2:1 resonant planets without geothermal heating. All planets are at 0.2 eccentricity. Results shown for planets orbiting stars from 2600 K (dark red line) to the Sun (blue line). Unfilled squares are temperatures encountered near the end of simulations that crashed due to a numeric instability but where the radiative imbalance is declining toward zero and is less than 5 W m$^{-2}$. (Bottom row) 1 mb specific humidity (g kg$^{-1}$) vs. S0X. Horizontal dotted line is plotted at the 3 g kg$^{-1}$ Kasting limit discussed in text.

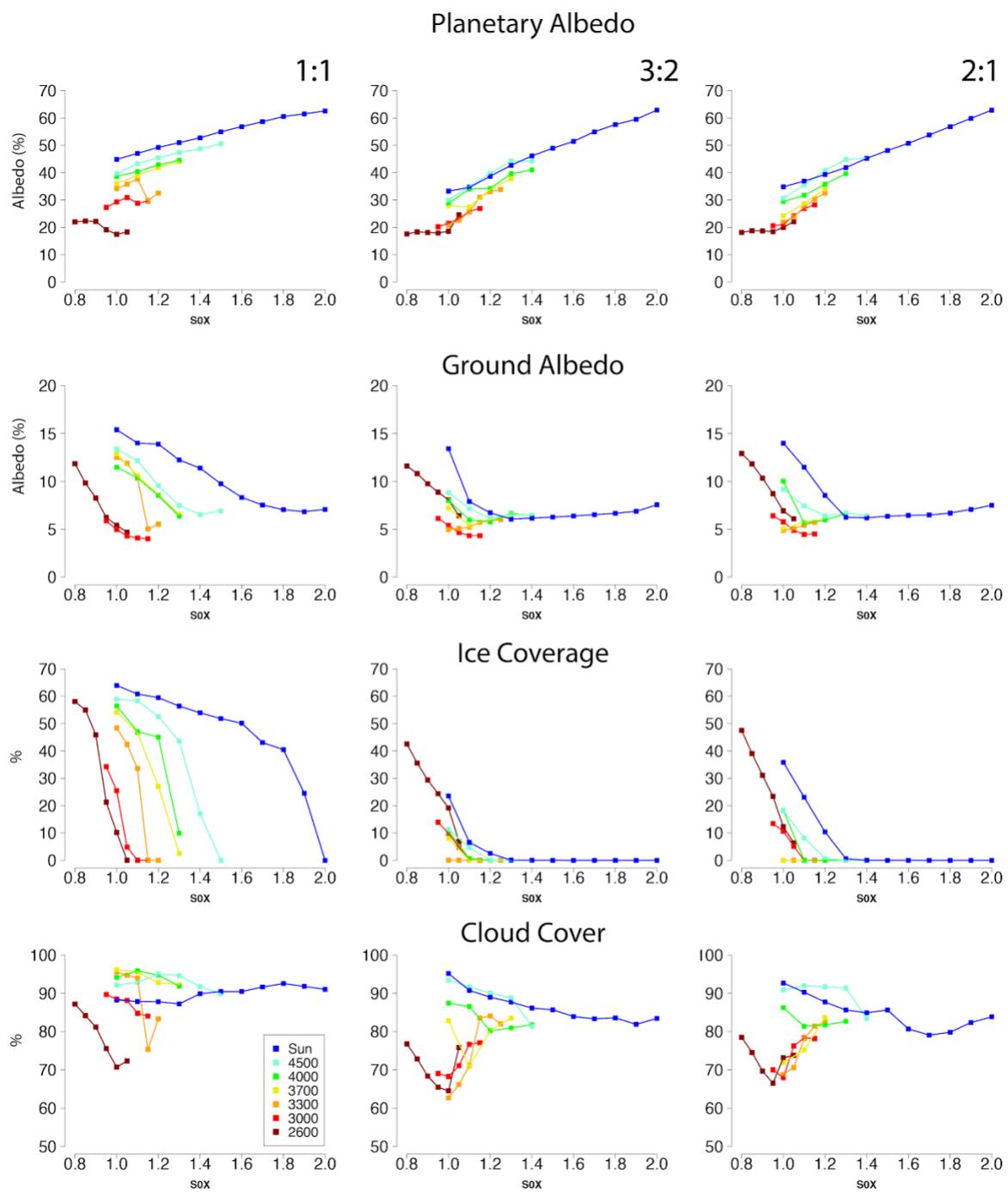

Figure 4. Select variables (all %) vs. S0X for (left) 1:1 (middle) 3:2 (right) 2:1 resonant planets without geothermal heating. (Top Row) Planetary Albedo, (Second Row) Ground Albedo (%), Planetary ice coverage, (bottom row) total cloud cover.

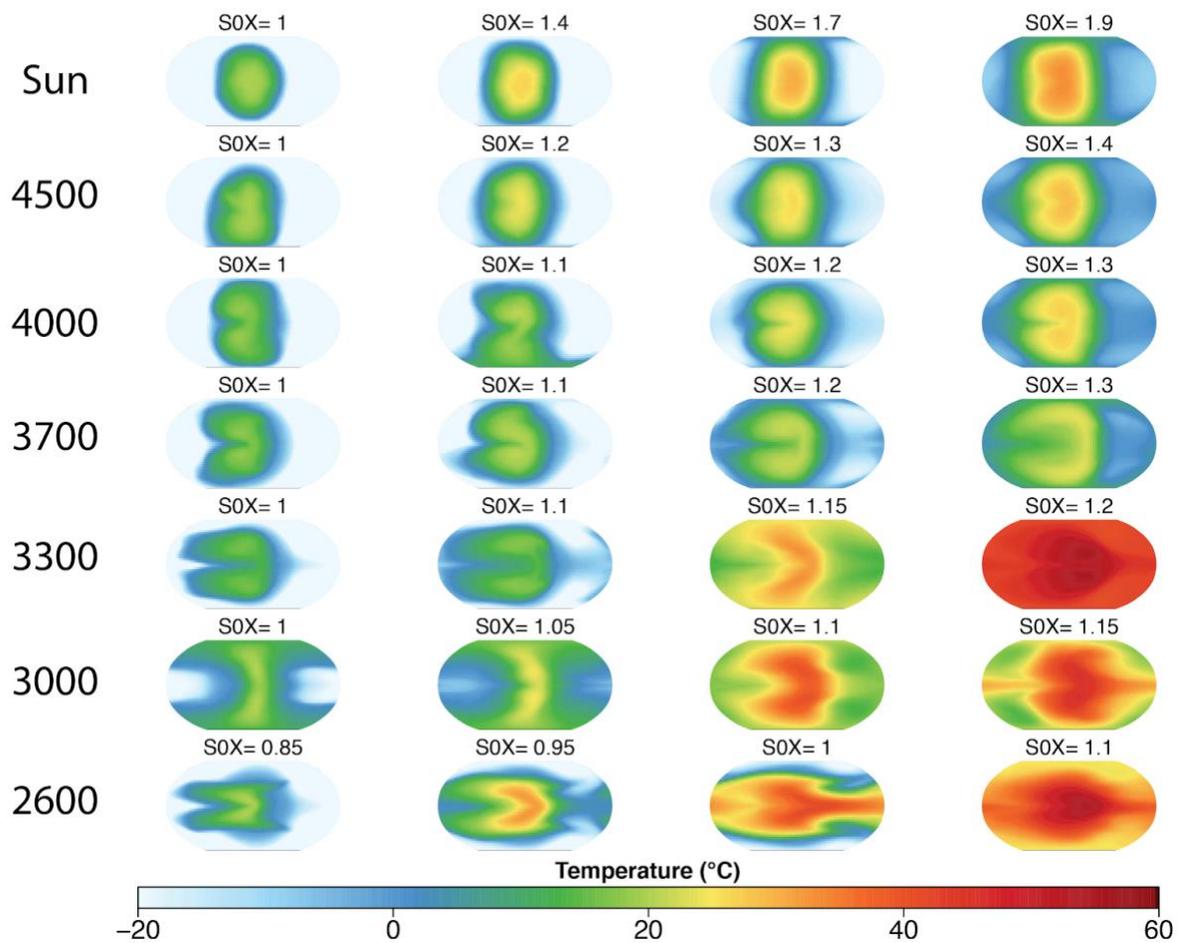

Figure 5. Surface air temperature field (°C) for selected 1:1 resonance cases of simulated planets orbiting a Sun-like star, and 4500 K, 4000 K, 3700 K, 3300 K, 3000 K, and 2600 K stars discussed in text. S0X increases from left to right for each star type. Note that the sampled S0X values may differ between rows.

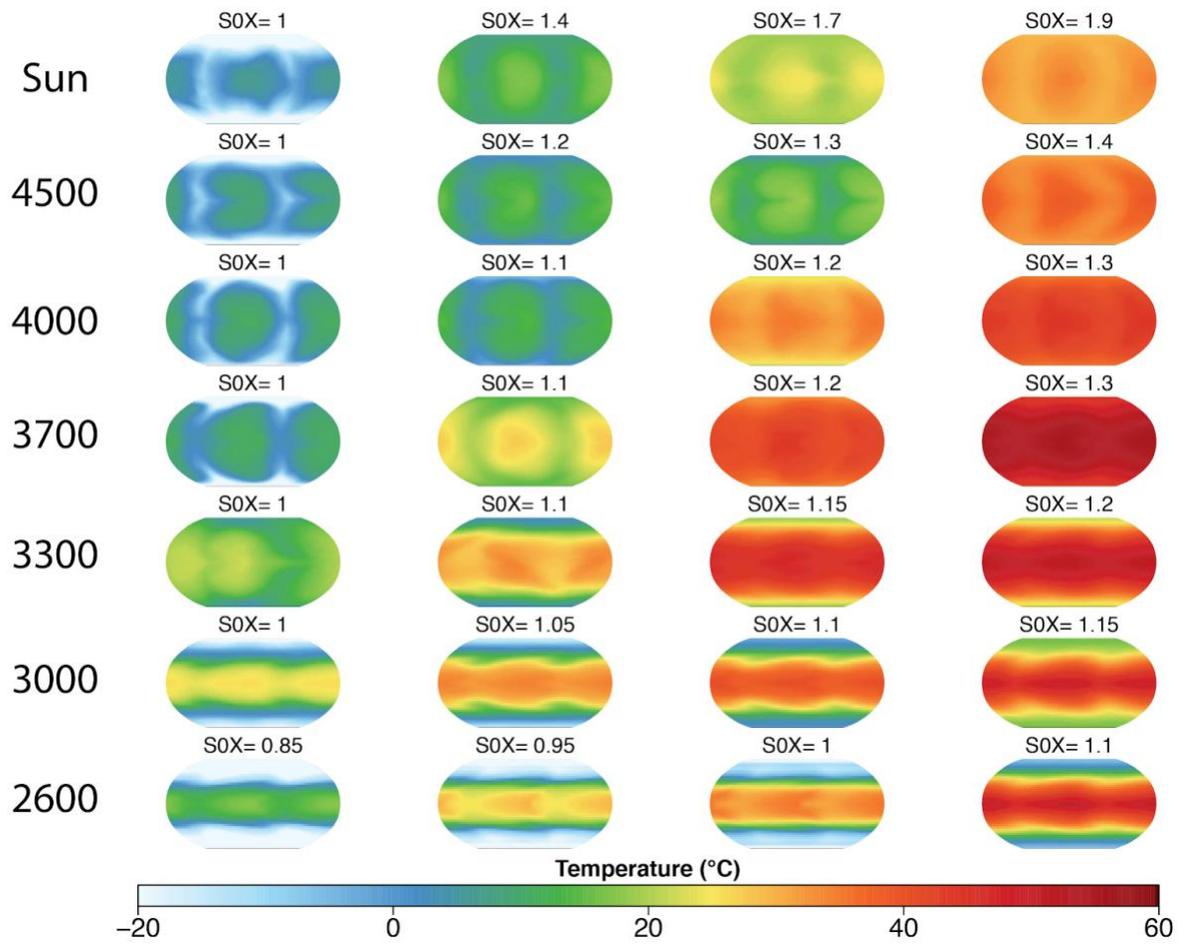

Figure 6. As in Figure 5, except the results are shown for 3:2 resonance planets.

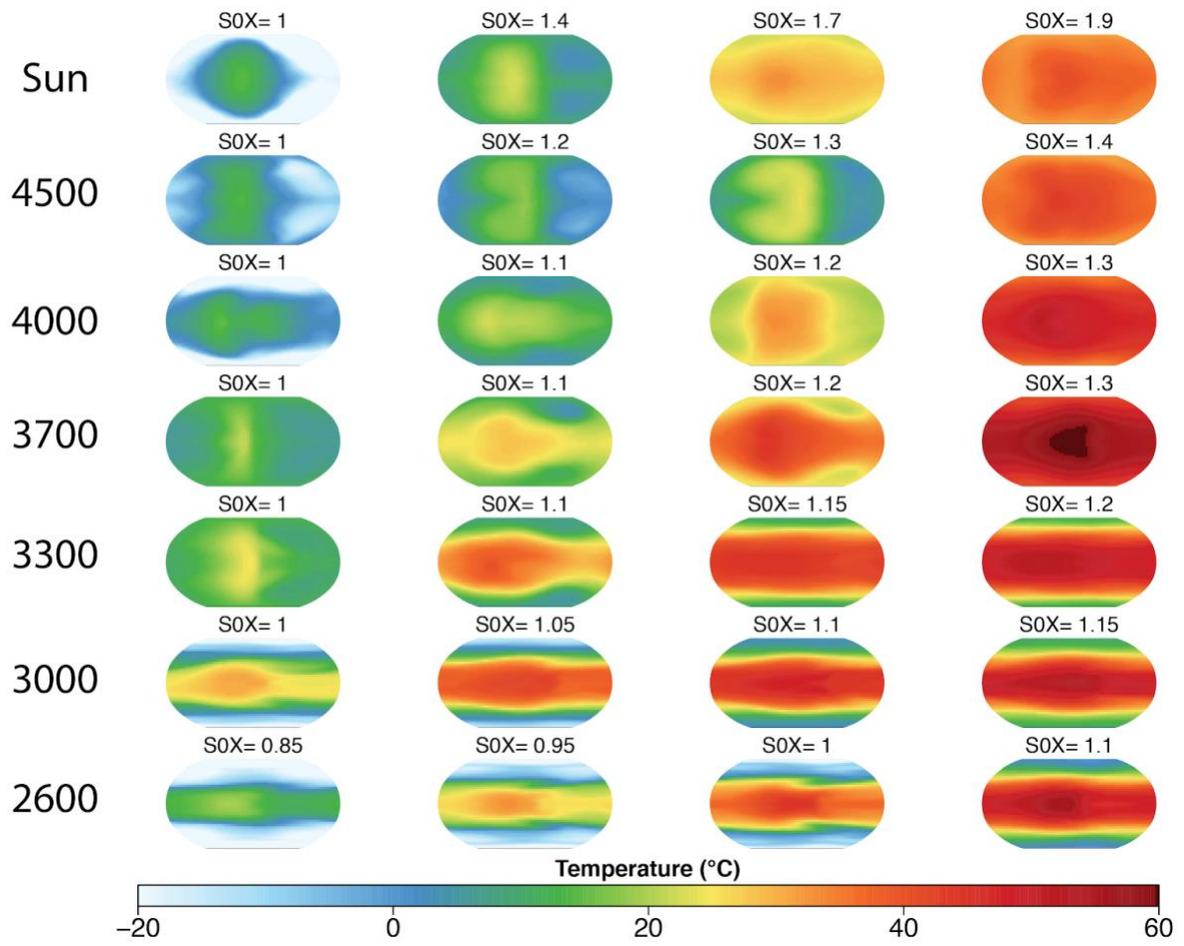

Figure 7. As in Figure 5-6, except the results are shown for 2:1 resonance planets.

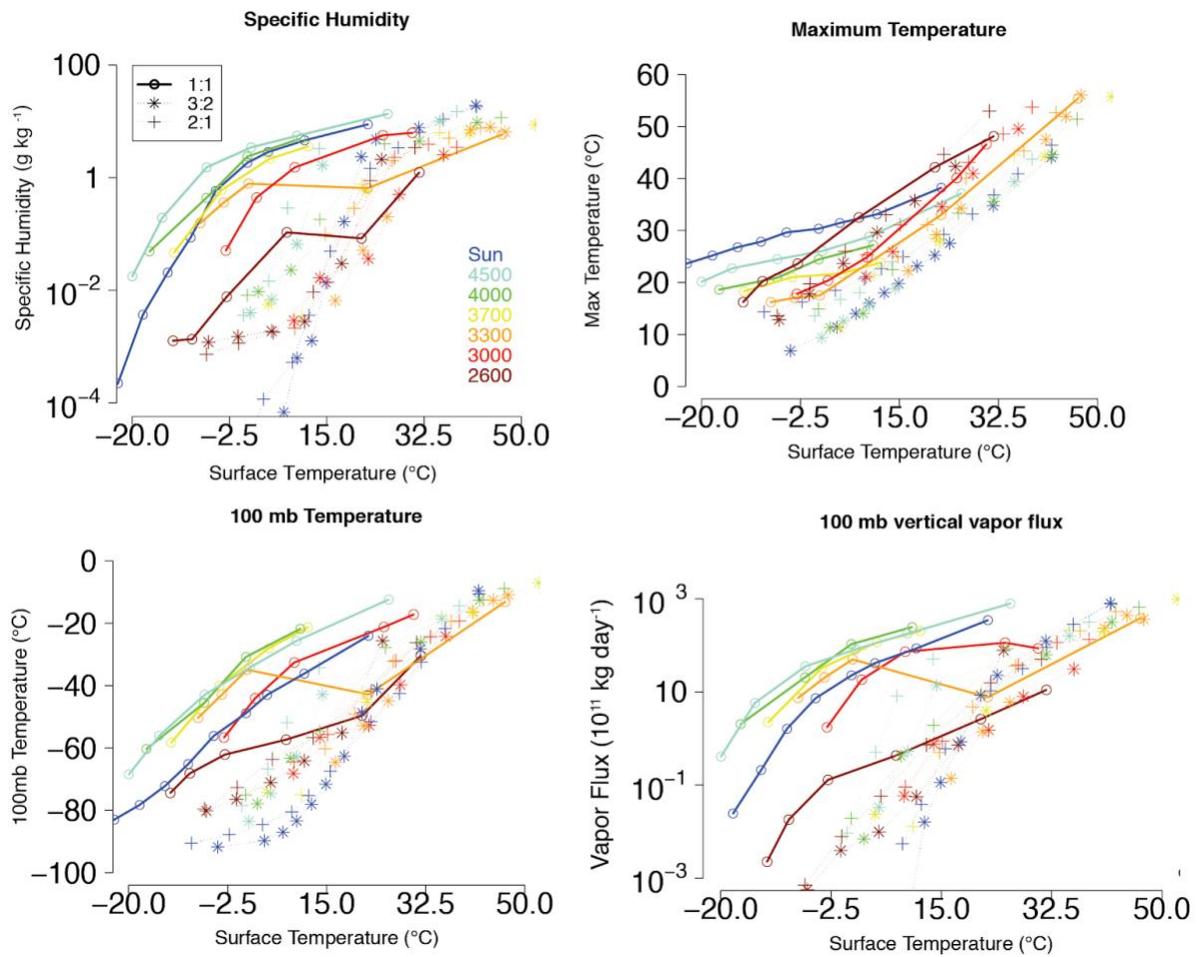

Figure 8. Selected variables plotted against global-mean surface temperature for different stellar types (colored as in figure 2) and (solid line) 1:1, or (scatter points) for 3:2 and 2:1 resonances. Shown is (top left) specific humidity (g kg$^{-1}$), (top right) maximum annual-mean temperature value (°C), (bottom left) 100 mb temperature (°C), (bottom right) total 100 mb vertical vapor flux ($10^{11}$ kg Earth day$^{-1}$).

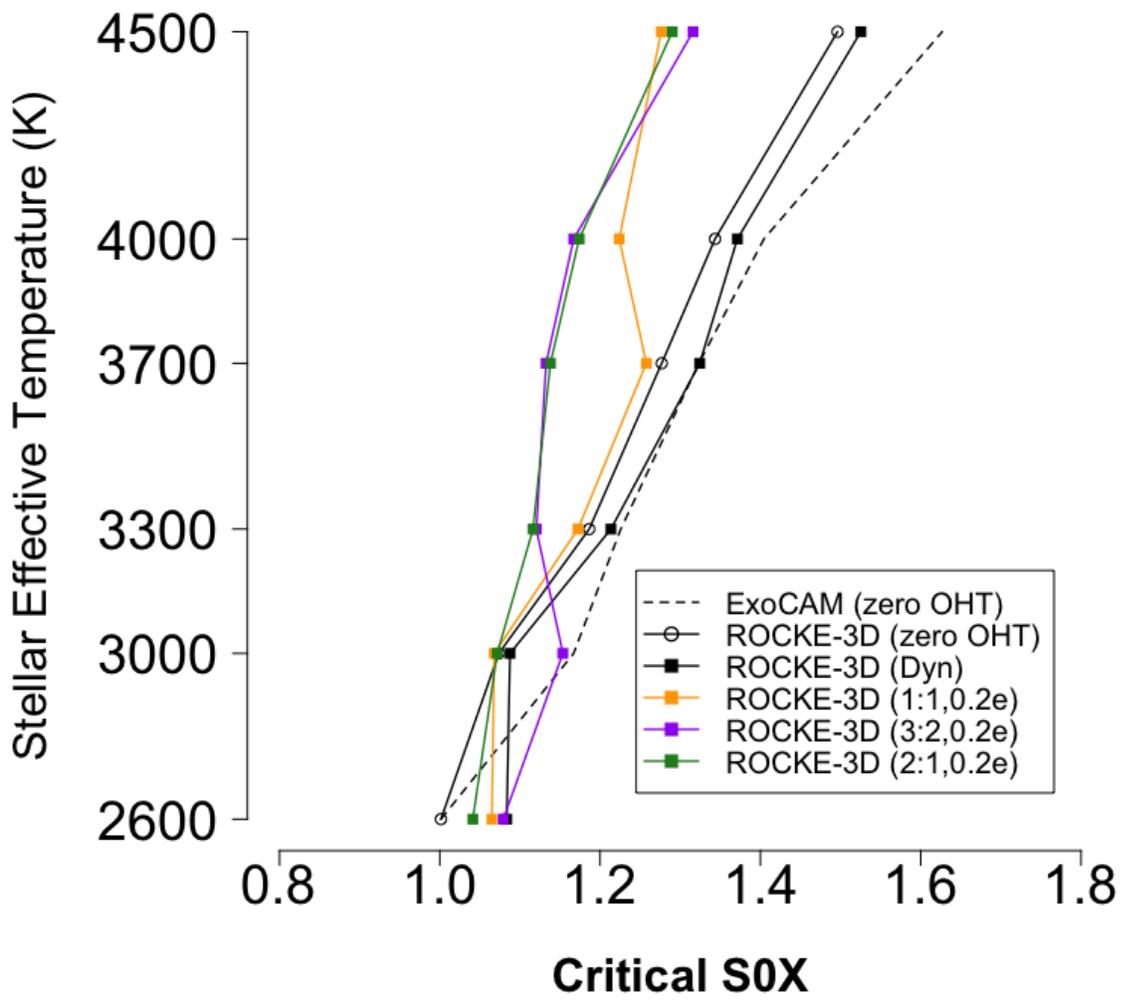

Figure 9. S0X value for which planets orbiting different spectral types first enter either a moist or runaway greenhouse regime. Results shown for synchronous rotation and zero eccentricity experiments from section (3.1) [exoCAM (dashed line), ROCKE-3d slab ocean (solid black line with open circles), ROCKE-3d 900 m dynamic ocean (solid black line with solid squares)] and section (3.2) [ROCKE-3d 158 m dynamic ocean runs at 0.2 eccentricity for 1:1 (orange), 3:2 (purple), or 2:1 resonance (green)].

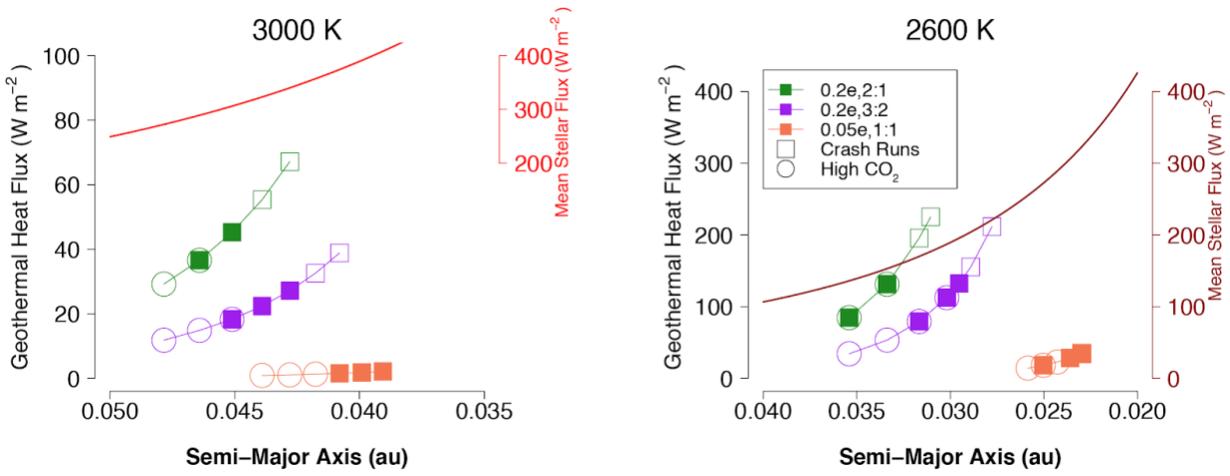

Figure 10. Geothermal heating (W m$^{-2}$) calculated from VPLANET vs. semi-major axis for the planets sampled orbiting the (left) 3000 K (right) 2600 K stars. Heating shown for 2:1 resonance, 0.2 eccentricity (green), 3:2 resonance, 0.2 eccentricity (purple), and 1:1, 0.05 eccentricity (orange) planets. Stellar masses of 0.143 M☉ (3000 K) and 0.086 M☉ (2600 K, as in Kopparapu et al., 2017) assumed for calculating the semi-major axis corresponding to a given stellar flux. Values where the corresponding ROCKE-3D simulation reached a stable equilibrium shown as a filled square, and crashed simulations shown as an open square. Runs performed with high $CO_2$ (1%) shown as open circles. Incident stellar flux at the semi-major axis distributed over a sphere [$S_0/4$] also shown for comparison. Note the difference in horizontal and vertical scales between the two plots.

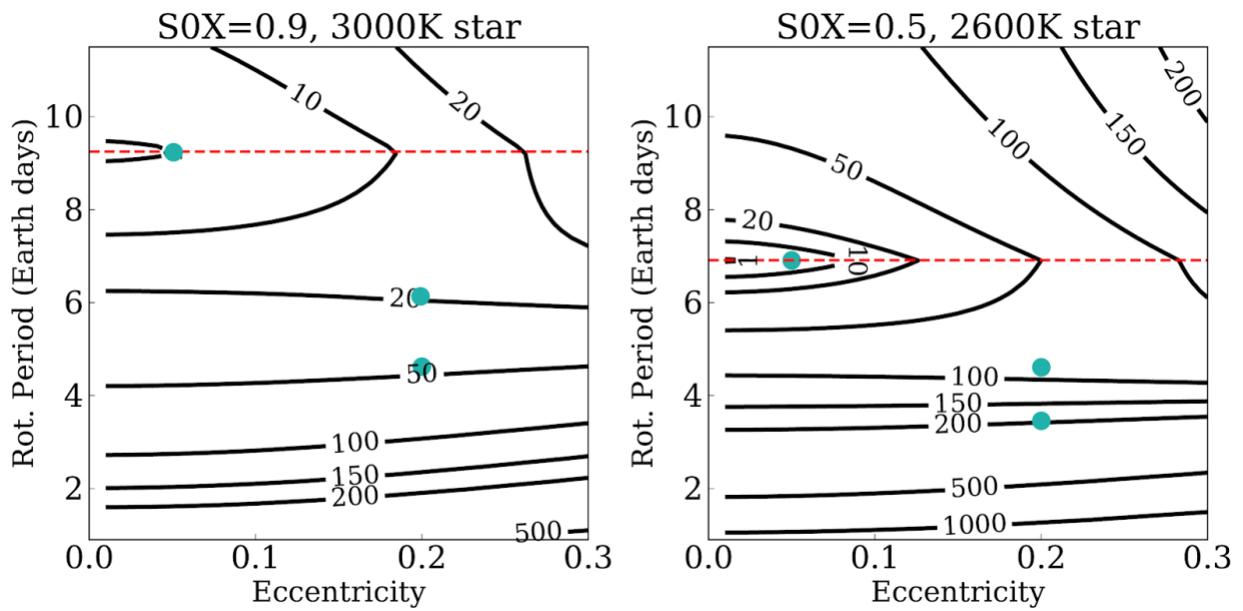

Figure 11. Contour map of tidal heating, *G* (W m$^{-2}$), for a rotation period and eccentricity phase space for two star-planet orbital configurations (left, S0X=0.9 for the 3000 K star; right, S0X=0.5, 2600 K star). Horizontal dashed red line is shown at the orbital period (1:1 resonance) and blue circles correspond to the four ROCKE-3D simulations performed within each phase space (0.05 and 0.2 eccentricity and for the 3:2 and 2:1 resonance). Calculations of tidal heating from VPLanet assume an Earth mass planet, stellar masses of 0.0886 M$_\odot$ and 0.143 M$_\odot$ for the 2600K and 3000K star, respectively (as in Kopparapu et al., 2017), a tidal Q factor of 100, and a Love number of degree 2 equal to 0.3.

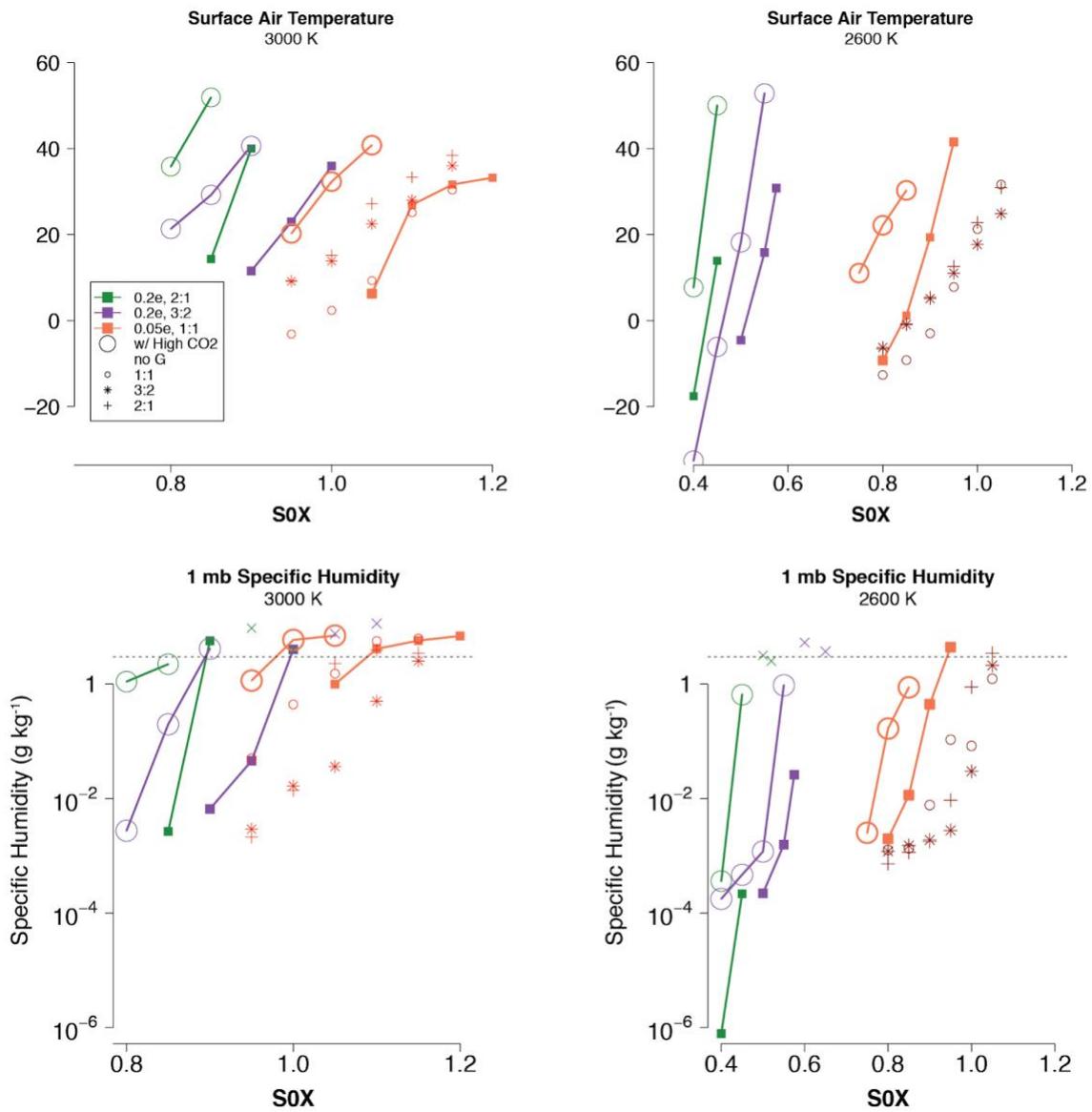

Figure 12. (Top row) Global-mean surface air temperature (°C) vs. S0X for runs with geothermal heating enabled for (left) 3000 K stars (right) 2600 K stars. Results shown for 3:2 (purple) or 2:1 (green) resonant planets at 0.2 eccentricity, and 1:1 (orange) resonant planets at 0.05 eccentricity. Lines connected by solid squares correspond to pure $N_2$ atmospheres and lines connected by larger unfilled circles correspond to simulations with 1% $CO_2$. Humidity values encountered near the end of crashed simulations are plotted as small x data points. Smaller symbols unconnected by lines (see legend) correspond to 0.2 eccentricity simulations without geothermal heating for comparison (as in Figure 3). (Bottom row) 1 mb specific humidity (g kg$^{-1}$) vs. S0X. Horizontal dotted line is plotted at the 3 g kg$^{-1}$ Kasting limit discussed in text.

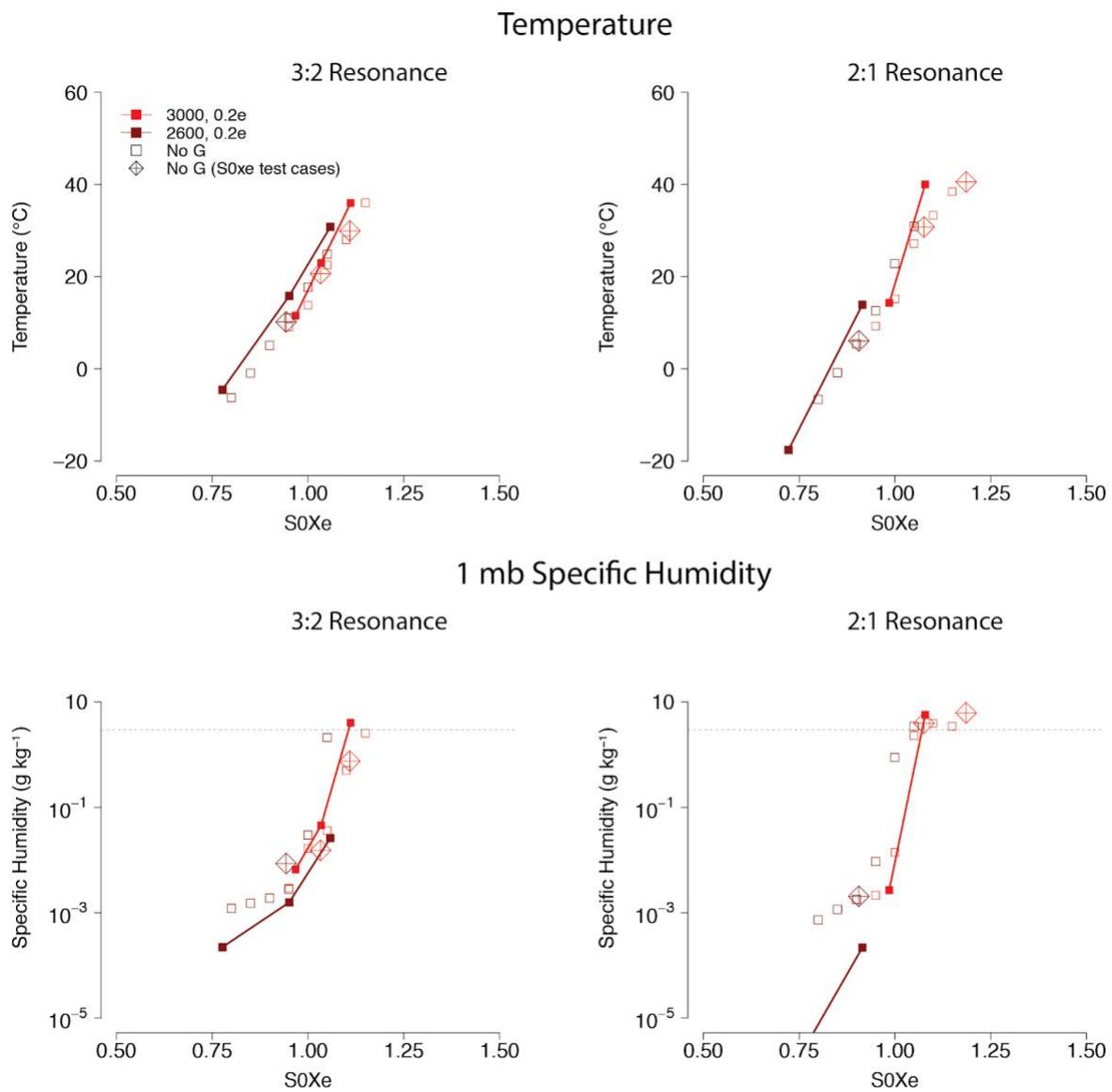

Figure 13. (Top) Global-mean surface air temperature (°C) for geothermal runs (solid lines) at (left) 3:2 and (right) 2:1 resonance plotted against S0Xe (see section 3.4). Zero geothermal runs (S0Xe = S0X) from section 3.2 are shown as small, unfilled squares. Zero geothermal simulations described in section 3.4 with deliberately chosen S0X to equal the S0Xe of the geothermal runs, and with the same rotation period, are shown as diamond-cross points.

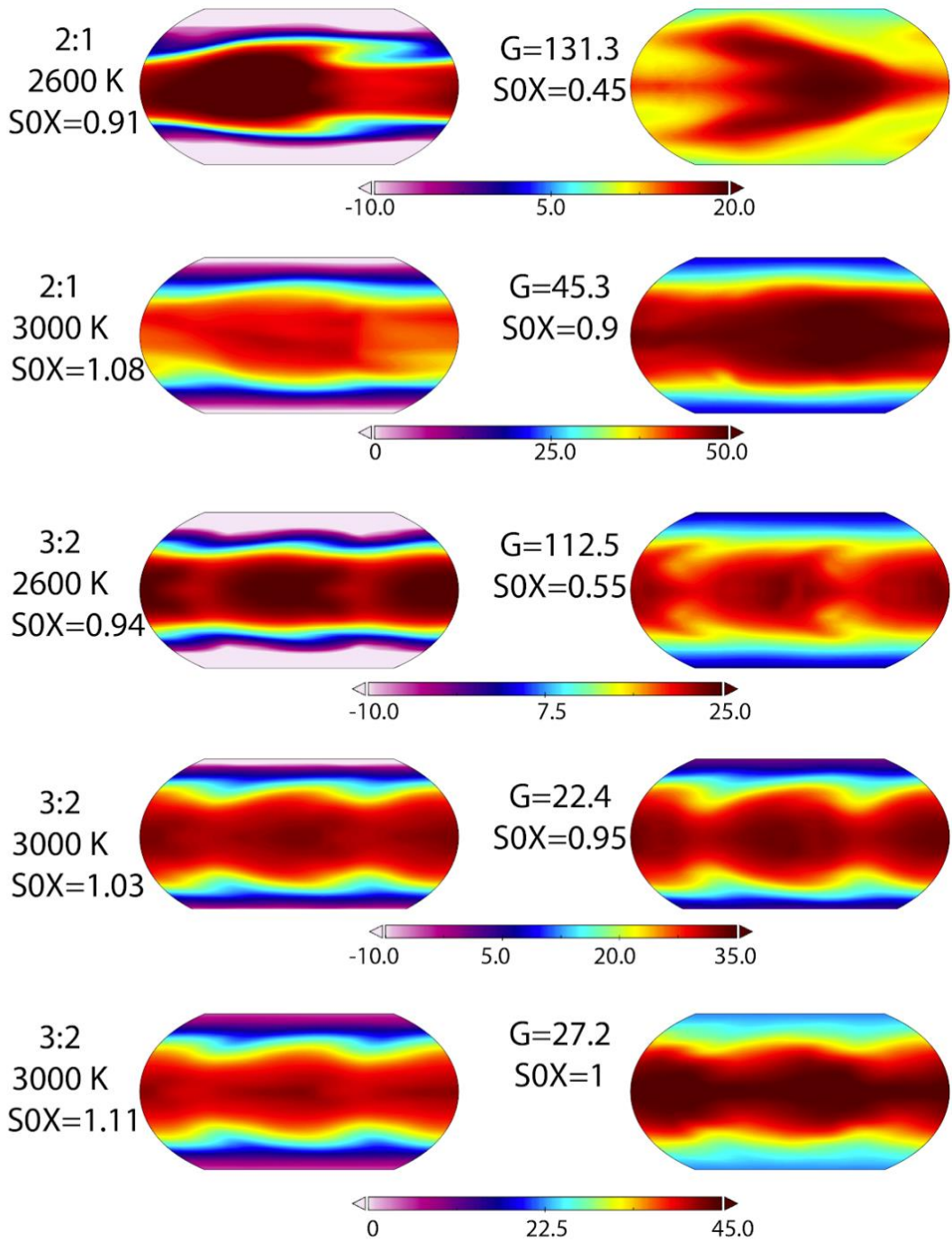

Figure 14. Five example comparisons of surface air temperature (°C) in which (along rows) each pair of planets features an identical S0Xe (as defined in text). Geothermal heated runs are shown on right, and no geothermal runs with elevated stellar flux shown on left. Values are shown for the resonance, star type, S0X, and geothermal heating (W m$^{-2}$). Results are for 3:2 or 2:1 resonant planets at 0.2 eccentricity. Note that the temperature scale differs for each planetary pair.

**Table 1.** Summary of runs in section 3.2

| Star | S0X | P (Earth Days) | Global-Mean $T_s$ (°C) 1:1 / 3:2 / 2:1 | 1 mb Specific Humidity (g kg$^{-1}$) 1:1 / 3:2 / 2:1 |
|---|---|---|---|---|
| Sun | 1.0 | 365.0 | -31.2 / -4.3 / -8.9 | 5.4·10$^{-6}$ / 2.7·10$^{-6}$ / 4.3·10$^{-6}$ |
| | 1.1 | 339.82 | -26.4 / 4.0 / -2.2 | 1.2·10$^{-4}$ / 8.1·10$^{-7}$ / 1.8·10$^{-6}$ |
| | 1.2 | 318.35 | -22.6 / 7.3 / 3.7 | 2.2·10$^{-4}$ / 6.9·10$^{-5}$ / 1.2·10$^{-4}$ |
| | 1.3 | 299.80 | -18.1 / 9.7 / 8.8 | 3.7·10$^{-3}$ / 6.3·10$^{-4}$ / 5.3·10$^{-4}$ |
| | 1.4 | 283.59 | -13.6 / 12.3 / 11.8 | 2.1·10$^{-2}$ / 1.3·10$^{-3}$ / 3.6·10$^{-3}$ |
| | 1.5 | 269.30 | -9.5 / 14.9 / 15.6 | 8.6·10$^{-2}$ / 1.4·10$^{-2}$ / 4.9·10$^{-2}$ |
| | 1.6 | 256.57 | -5.0 / 18.1 / 22.8 | 0.6 / 0.2 / 1.5 |
| | 1.7 | 245.16 | 0.7 / 21.2 / 27.9 | 1.8 / 2.4 / 3.4 |
| | 1.8 | 234.88 | 4.5 / 23.9 / 31.7 | 2.9 / 4.8 / 6.4 |
| | 1.9 | 225.54 | 11.0 / 31.6 / 36.0 | 4.6 / 7.8 / 10.9 |
| | 2.0 | 217.03 | 22.4 / 41.8 / 41.9 | 8.8 / 19.3 / 18.2 |
| 4500 | 1.0 | 123.30 | -20.0 / 1.3 / 0.0 | 1.8·10$^{-2}$ / 4.0·10$^{-3}$ / 3.6·10$^{-3}$ |
| | 1.1 | 114.80 | -14.6 / 5.1 / 4.6 | 1.9·10$^{-1}$ / 6.9·10$^{-3}$ / 1.5·10$^{-2}$ |
| | 1.2 | 107.54 | -6.6 / 9.7 / 8.0 | 1.5 / 6.6·10$^{-2}$ / 0.3 |
| | 1.3 | 101.28 | 1.4 / 14.2 / 13.7 | 3.5 / 1.7 / 3.3 |
| | 1.4 | 95.80 | 9.7 / 35.3 / 38.4 | 5.5 / 10.1 / 14.9 |
| | 1.5 | 90.97 | 26.0 / 89.4** / 99.5** | 13.6 / 252.1** / 386.3** |
| 4000 | 1.0 | 74.29 | -16.9 / 2.7 / 0.7 | 4.9·10$^{-2}$ / 9.5·10$^{-3}$ / 8.2·10$^{-3}$ |
| | 1.1 | 69.17 | -6.7 / 8.6 / 13.8 | 0.44 / 2.3·10$^{-2}$ / 0.18 |
| | 1.2 | 64.80 | 0.7 / 31.6 / 25.4 | 2.4 / 4.5 / 4.0 |
| | 1.3 | 61.02 | 10.4 / 42.0 / 46.4 | 5.0 / 9.6 / 11.7 |
| | 1.4 | 57.72 | 57.7* / 61.7* / 54.5** | 7.2* / 7.9* / 11.6** |
| 3700 | 1.0 | 47.71 | -12.5 / 4.5 / 10.5 | 4.6·10$^{-2}$ / 5.7·10$^{-3}$ / 3.4·10$^{-3}$ |
| | 1.1 | 44.42 | -4.0 / 22.1 / 21.3 | 0.6 / 0.6 / 0.8 |
| | 1.15 | 42.96 | 1.0 / 36.7 / 27.7 | 1.3 / 4.3 / 3.7 |
| | 1.2 | 41.61 | 4.8 / 40.7 / 35.3 | 2.2 / 6.2 / 6.3 |
| | 1.3 | 39.19 | 11.7 / 52.5 / 53.9* | 3.6 / 8.8 / 4.4* |
| | 1.4 | 37.07 | 43.6** / 55.7** / 54.5** | 7.9 / 10.7** / 7.3** |
| 3300 | 1.0 | 22.64 | -7.7 / 16.5 / 14.7 | 0.2 / 6.6·10$^{-3}$ / 0.1 |
| | 1.05 | 21.83 | -3.6 / 21.6 / 20.0 | 0.4 / 0.1 / 0.3 |
| | 1.1 | 21.08 | 0.9 / 25.9 / 26.8 | 0.8 / 0.2 / 2.0 |
| | 1.15 | 20.39 | 22.4 / 40.9 / 37.0 | 0.7 / 7.0 / 5.1 |
| | 1.2 | 19.75 | 46.6 / 44.4 / 42.7 | 5.9 / 7.9 / 7.6 |
| | 1.25 | 19.15 | 50.6* / 47.1 / 44.0 | 9.3 / 6.5 / 5.8 |
| 3000 | 0.95 | 8.88 | -3.2 / 9.1 / 9.3 | 0.1 / 2.9·10$^{-3}$ / 2.1·10$^{-3}$ |
| | 1.0 | 8.54 | 2.4 / 13.8 / 15.1 | 0.4 / 1.7·10$^{-2}$ / 1.4·10$^{-2}$ |
| | 1.05 | 8.23 | 9.3 / 22.5 / 27.2 | 1.52 / 3.7·10$^{-2}$ / 2.3 |
| | 1.1 | 7.95 | 25.1 / 28.0 / 33.2 | 5.7 / 0.5 / 3.9 |
| | 1.15 | 7.69 | 30.4 / 36.0 / 38.4 | 6.3 / 2.6 / 3.5 |
| | 1.2 | 7.45 | 38.9* / 44.2** / 41.7** | 3.6* / 8.8** / 7.6** |
| 2600 | 0.8 | 4.85 | -12.7 / -6.3 / -6.6 | 1.3·10$^{-3}$ / 1.2·10$^{-3}$ / 7.3·10$^{-4}$ |
| | 0.85 | 4.64 | -9.2 / -1.0 / -0.8 | 1.4·10$^{-3}$ / 1.5·10$^{-3}$ / 1.2·10$^{-3}$ |
| | 0.9 | 4.44 | -3.0 / 5.1 / 5.4 | 7.7·10$^{-3}$ / 1.9·10$^{-3}$ / 1.8·10$^{-3}$ |
| | 0.95 | 4.27 | 7.8 / 11.0 / 12.6 | 0.1 / 2.8·10$^{-3}$ / 9.4·10$^{-3}$ |
| | 1.0 | 4.10 | 21.3 / 17.7 / 22.8 | 0.1 / 3·10$^{-2}$ / 0.9 |
| | 1.05 | 3.96 | 31.7 / 24.9 / 30.9 | 1.2 / 2.1 / 3.5 |
| | 1.1 | 3.82 | 37.7* / 32.4* / 37.3* | 7.1 / 3.6* / 3.8* |

Table 1. List of climate simulations performed with ROCKE-3D at 0.2 eccentricity and with no geothermal heating. Shown is stellar type, S0X, orbital period, global-mean temperature (°C), and 1 mb specific humidity (g kg$^{-1}$) for the 1:1, 3:2, and 2:1 resonances. Data points with a single star (*) represent runs that crashed due to a numeric instability but with a radiative imbalance less than 5 W m$^{-2}$ that was declining with time. Double stars indicate simulations with a crash and imbalance greater than 5 W m$^{-2}$.

**Table 2.** Summary of runs in section 3.3

| Star | S0X | $e$ | P (Earth Days) | $G$ (W m$^{-2}$) | Global-Mean T$_s$ (°C) | 1 mb Specific Humidity (g kg$^{-1}$) |
|---|---|---|---|---|---|---|
| 3000 | **1:1 Resonance** | | | | | |
| | 0.95 | 0.05 | 8.88 | 0.89 | [20.3] | [1.2] |
| | 1.0 | 0.05 | 8.54 | 1.08 | [32.3] | [5.9] |
| | 1.05 | 0.05 | 8.23 | 1.30 | 6.3 [40.8] | 1.0 [7.0] |
| | 1.1 | 0.05 | 7.95 | 1.55 | 26.9 | 4.1 |
| | 1.15 | 0.05 | 7.69 | 1.83 | 31.6 | 5.8 |
| | **3:2 Resonance** | | | | | |
| | 0.8 | 0.2 | 10.09 | 11.83 | [21.3] | [2.7·10$^{-3}$] |
| | 0.85 | 0.2 | 9.65 | 14.86 | [29.2] | [0.2] |
| | 0.9 | 0.2 | 9.24 | 18.30 | 11.5 [40.6] | 6.6·10$^{-3}$ [4.2] |
| | 0.95 | 0.2 | 8.88 | 22.40 | 23.0 [41.3**] | 4.6·10$^{-2}$ [1.7**] |
| | 1.0 | 0.2 | 8.54 | 27.18 | 36.0 | 4.0 |
| | 1.05 | 0.2 | 8.23 | 32.60 | 47.9** | 7.6** |
| | **2:1 Resonance** | | | | | |
| | 0.8 | 0.2 | 10.09 | 29.2 | [35.8] | [1.1] |
| | 0.85 | 0.2 | 9.65 | 36.58 | 14.3 [51.9] | 2.7·10$^{-3}$ [2.2] |
| | 0.9 | 0.2 | 9.24 | 45.27 | 40.0 [46.7**] | 5.7 [1.3**] |
| | 0.95 | 0.2 | 8.88 | 55.32 | 64.5** | 9.5** |
| 2600 | **1:1 Resonance** | | | | | |
| | 0.75 | 0.05 | 5.10 | 14.3 | [11.1] | [2.5·10$^{-3}$] |
| | 0.8 | 0.05 | 4.85 | 18.21 | -9.3 [22.2] | 2.0·10$^{-3}$ [0.2] |
| | 0.85 | 0.05 | 4.64 | 22.87 | 1.1 [30.3] | 0.01 [0.9] |
| | 0.9 | 0.05 | 4.44 | 28.3 | 19.3 [37.0**] | 0.44 [1.4**] |
| | 0.95 | 0.05 | 4.27 | 34.7 | 41.6 | 4.4 |
| | **3:2 Resonance** | | | | | |
| | 0.4 | 0.2 | 8.16 | 34.3 | [-32.6] | [1.8·10$^{-4}$] |
| | 0.45 | 0.2 | 7.47 | 53.3 | [-6.1] | [4.7·10$^{-4}$] |
| | 0.5 | 0.2 | 6.9 | 79.32 | -4.5 [18.2] | 2.2·10$^{-4}$ [1.2·10$^{-3}$] |
| | 0.55 | 0.2 | 6.43 | 112.5 | 15.8 [52.8] | 1.6·10$^{-3}$ [1.0] |
| | 0.575 | 0.2 | 6.22 | 132.4 | 30.8 [58.1**] | 0.03 [1.2**] |
| | 0.6 | 0.2 | 6.02 | 155.3 | 69.0** | 5.3** |
| | **2:1 Resonance** | | | | | |
| | 0.4 | 0.2 | 8.16 | 84.68 | -17.6 [7.7] | 7.8·10$^{-7}$ [3.6·10$^{-4}$] |
| | 0.45 | 0.2 | 7.47 | 131.25 | 13.9 [50.0] | 2.2·10$^{-4}$ [0.7] |
| | 0.5 | 0.2 | 6.91 | 195.6 | 67.8** | 3.2** |

Table 2. List of climate simulations performed with ROCKE-3D at 0.05 or 0.2 eccentricity and with geothermal heating enabled. Shown is stellar type (2600 K or 3000 K), eccentricity, orbital period, prescribed geothermal heat flux derived from VPLanet, temperature (°C), and 1 mb specific humidity (g kg$^{-1}$) for the 1:1, 3:2, and 2:1 resonances. Bracketed values indicate runs with 1% $CO_2$. Data points with a single star (*) represent runs that crashed due to a numeric instability but with a radiative imbalance less than 5 W m$^{-2}$ that was declining with time. Double stars indicate simulations with a crash and imbalance greater than 5 W m$^{-2}$.

Supplementary Video 1. Time evolution of incoming stellar flux (W m$^{-2}$) during one orbit (shown as one frame per model "month" where the duration of each frame is weighted by the length of the month) for a 1:1 resonance planet at 0.2 eccentricity. Grid lines correspond to the horizontal resolution of ROCKE-3D simulations in this paper.

Supplementary Video 2. As in Supplementary Video 1, except for a 2:1 resonance planet.

Supplementary Video 3. As in Supplementary Video 1-2, except for a 3:2 resonance planet.

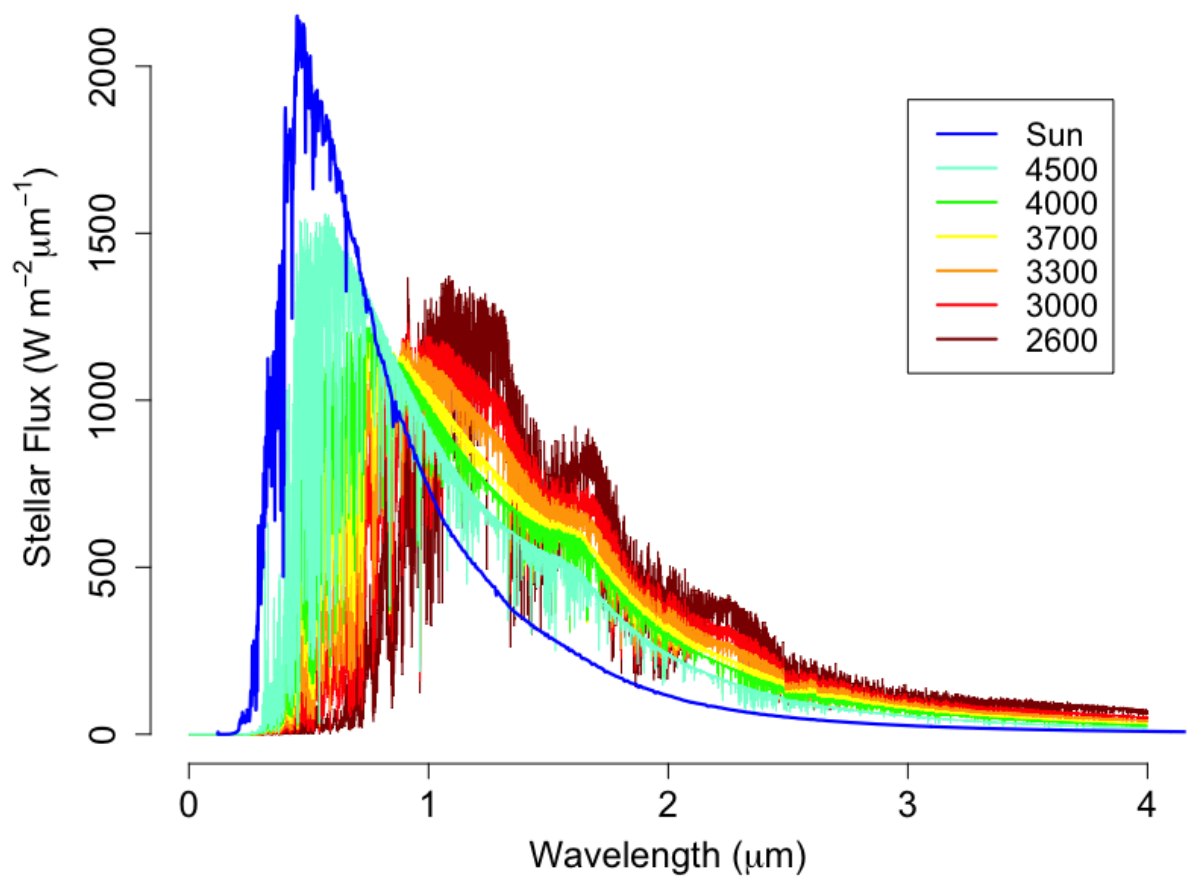

Figure S1. Stellar spectra for stars studied in this paper, based on the BT-SETTL model grid of theoretical spectra.

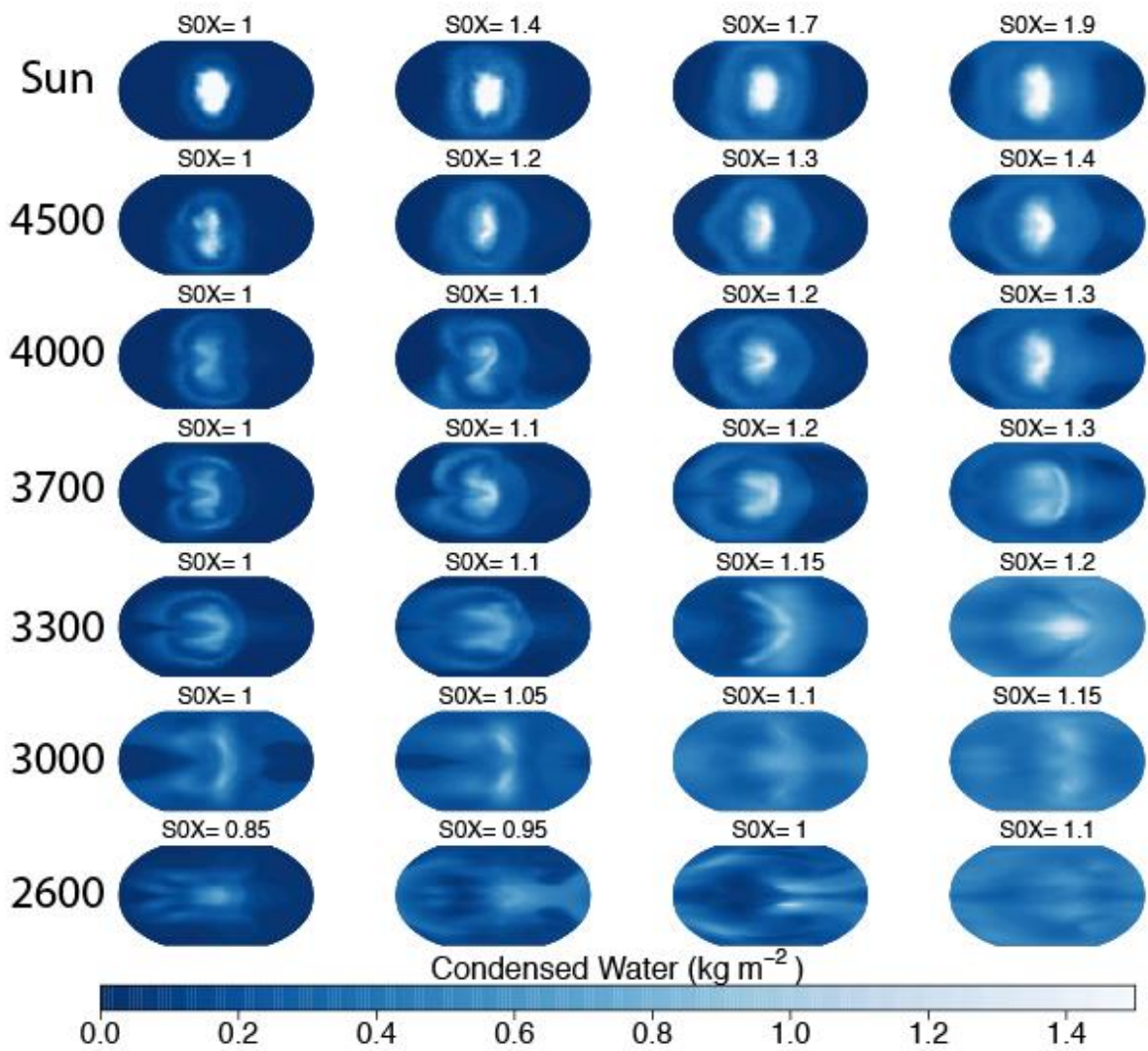

Figure S2. Cloud condensed water (kg m$^{-2}$) for select values of S0X (across rows) for different stellar types (columns). Results shown for 1:1 resonance planets.

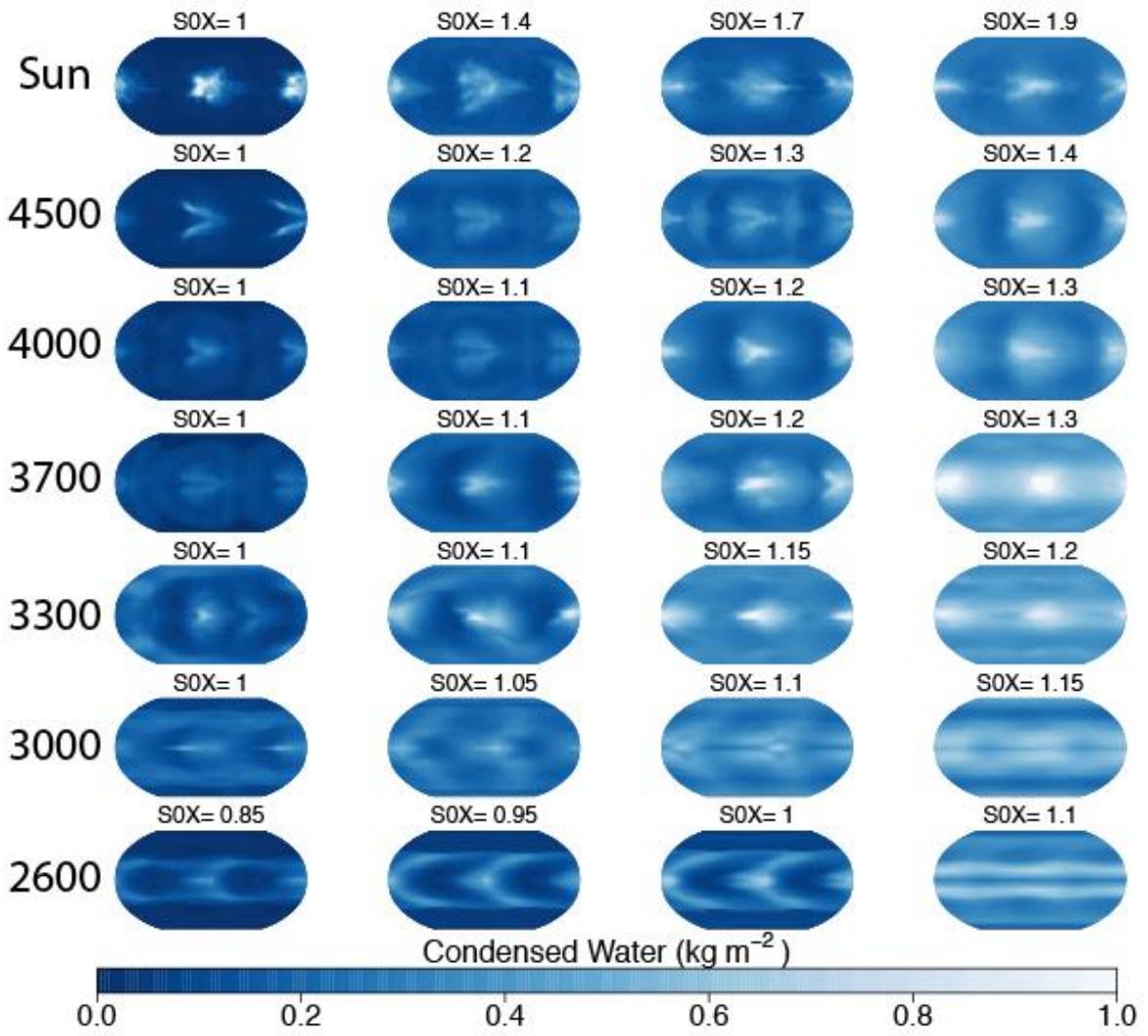

Figure S3. As in Figure S2, except results shown for 3:2 resonance planets.

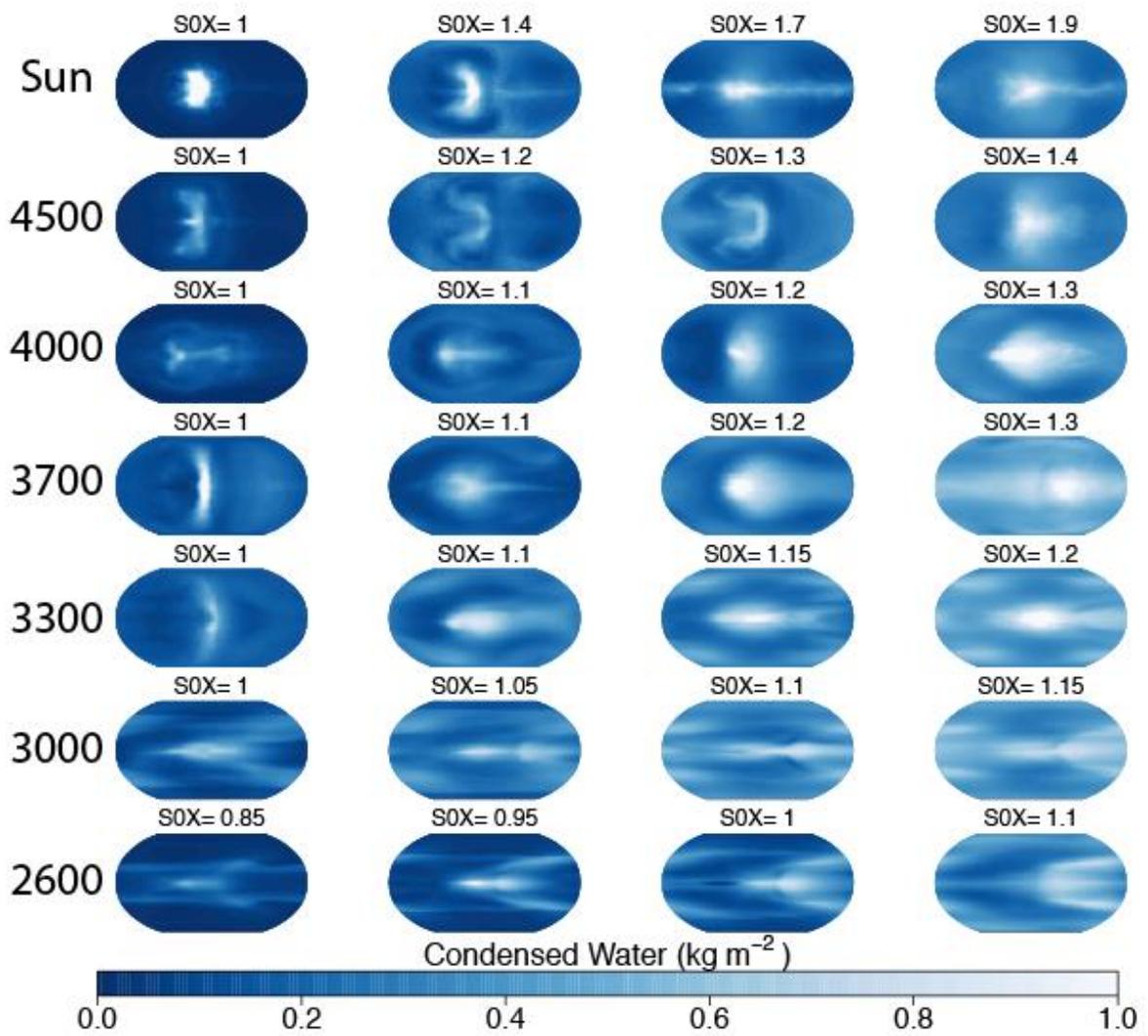

Figure S4. As in Figure S2-S3, except results shown for 2:1 resonance planets.

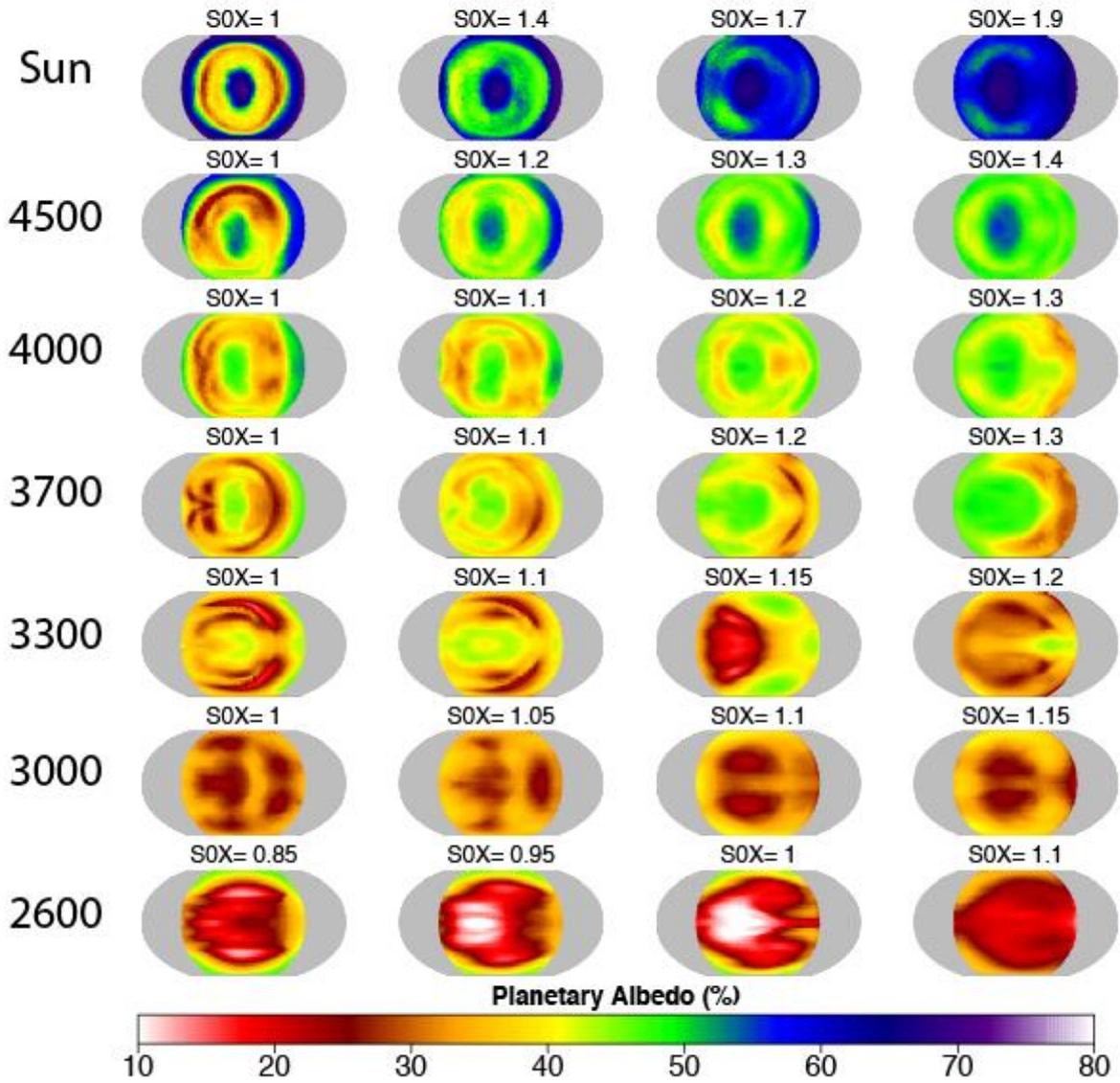

Figure S5. Planetary albedo (%) for select values of S0X (across rows) for different stellar types (columns). Results shown for 1:1 resonance planets.

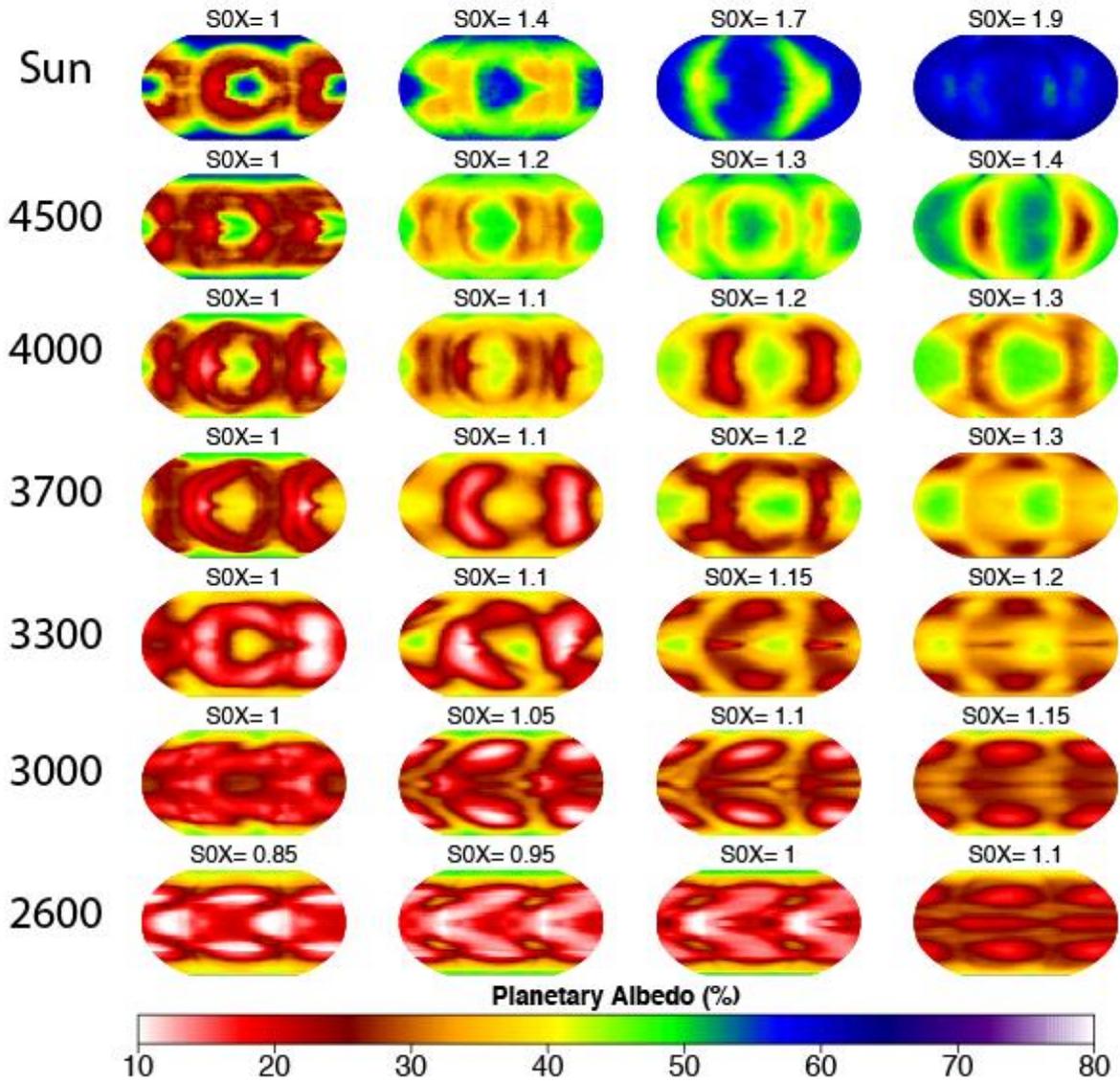

Figure S6. As in Figure S5, except results shown for 3:2 resonance planets.

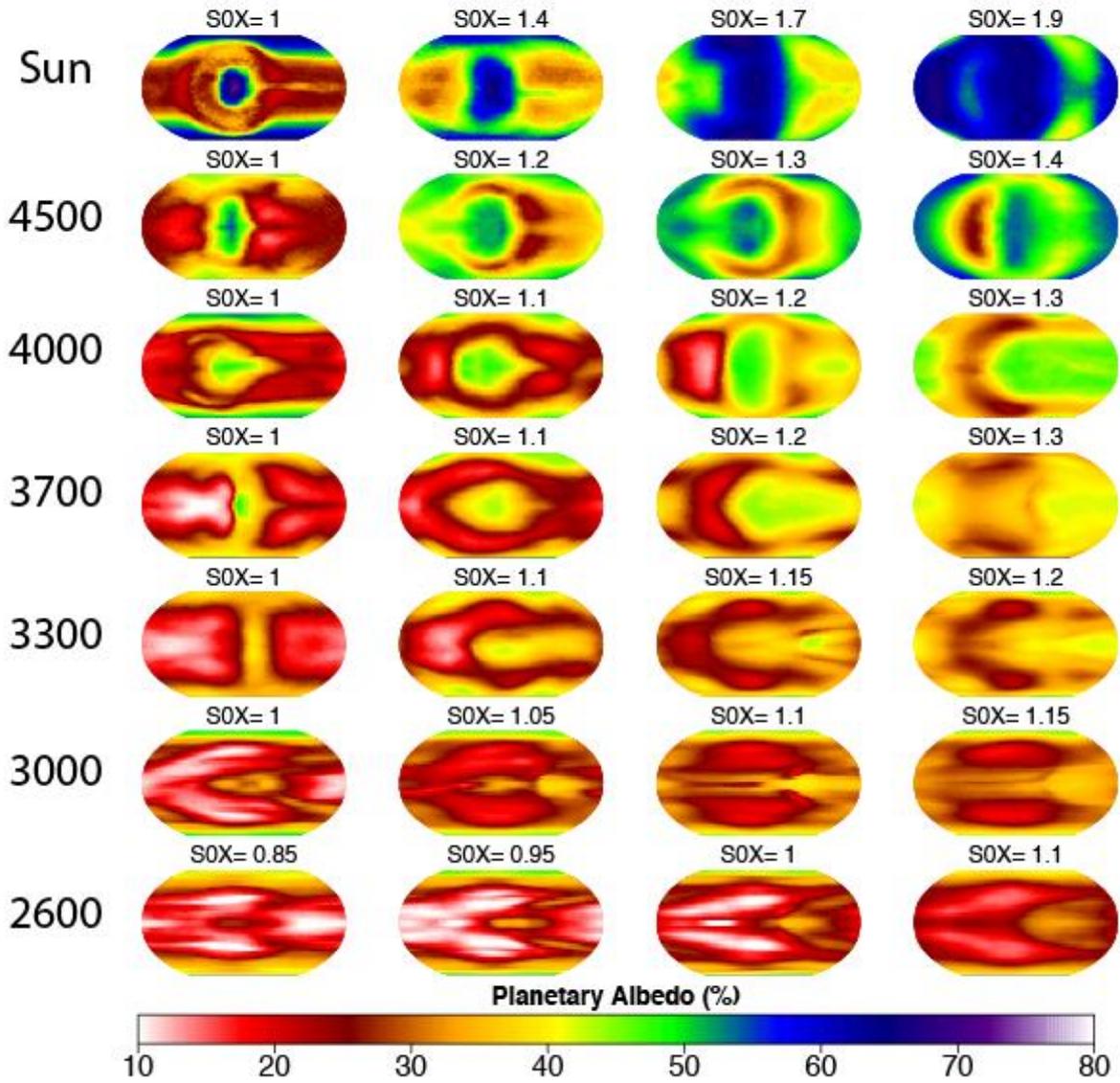

Figure S7. As in Figure S5-S6, except results shown for 2:1 resonance planets.

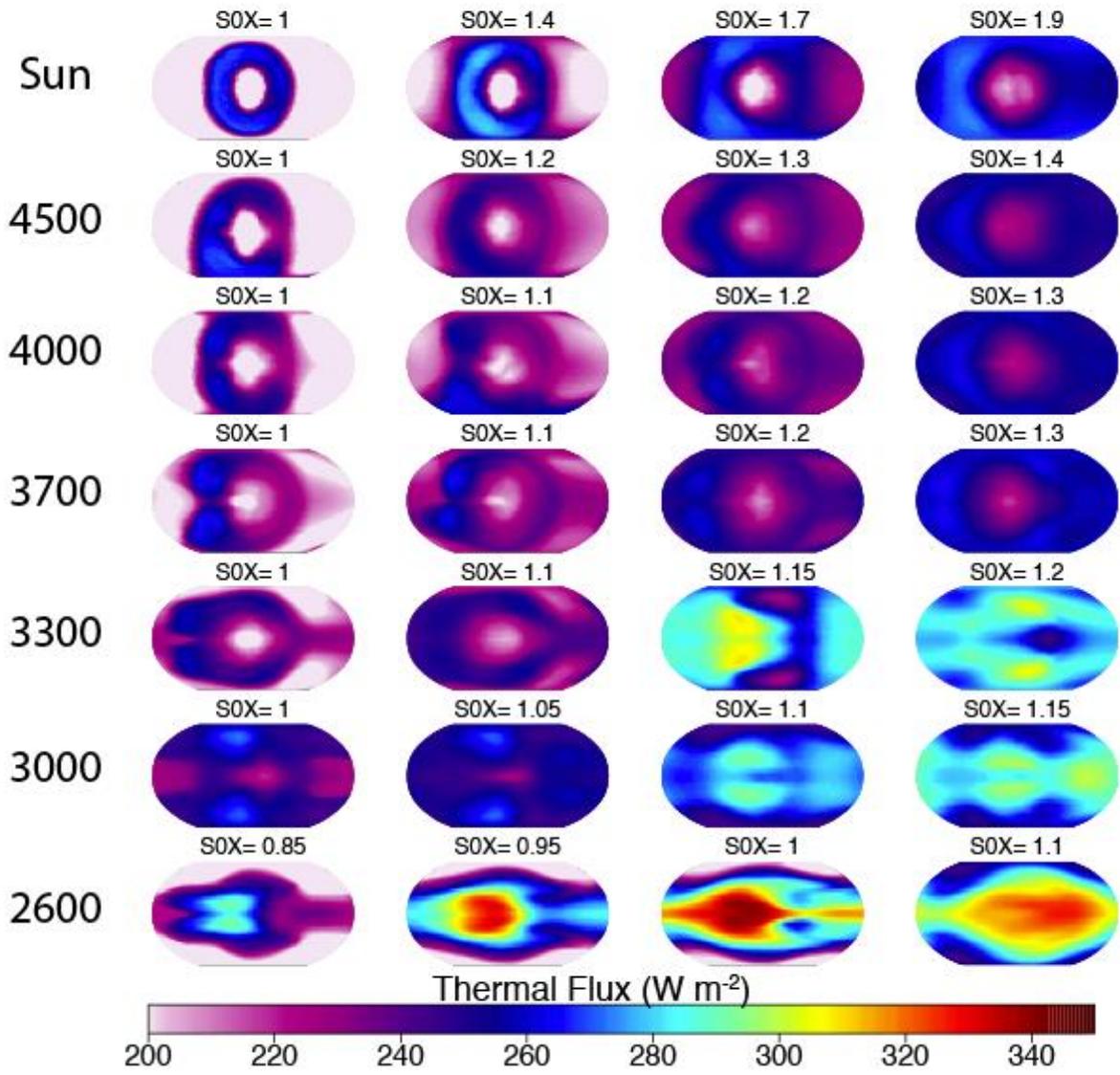

Figure S8. Outgoing longwave radiation (W m$^{-2}$) for select values of S0X (across rows) for different stellar types (columns). Results shown for 1:1 resonance planets.

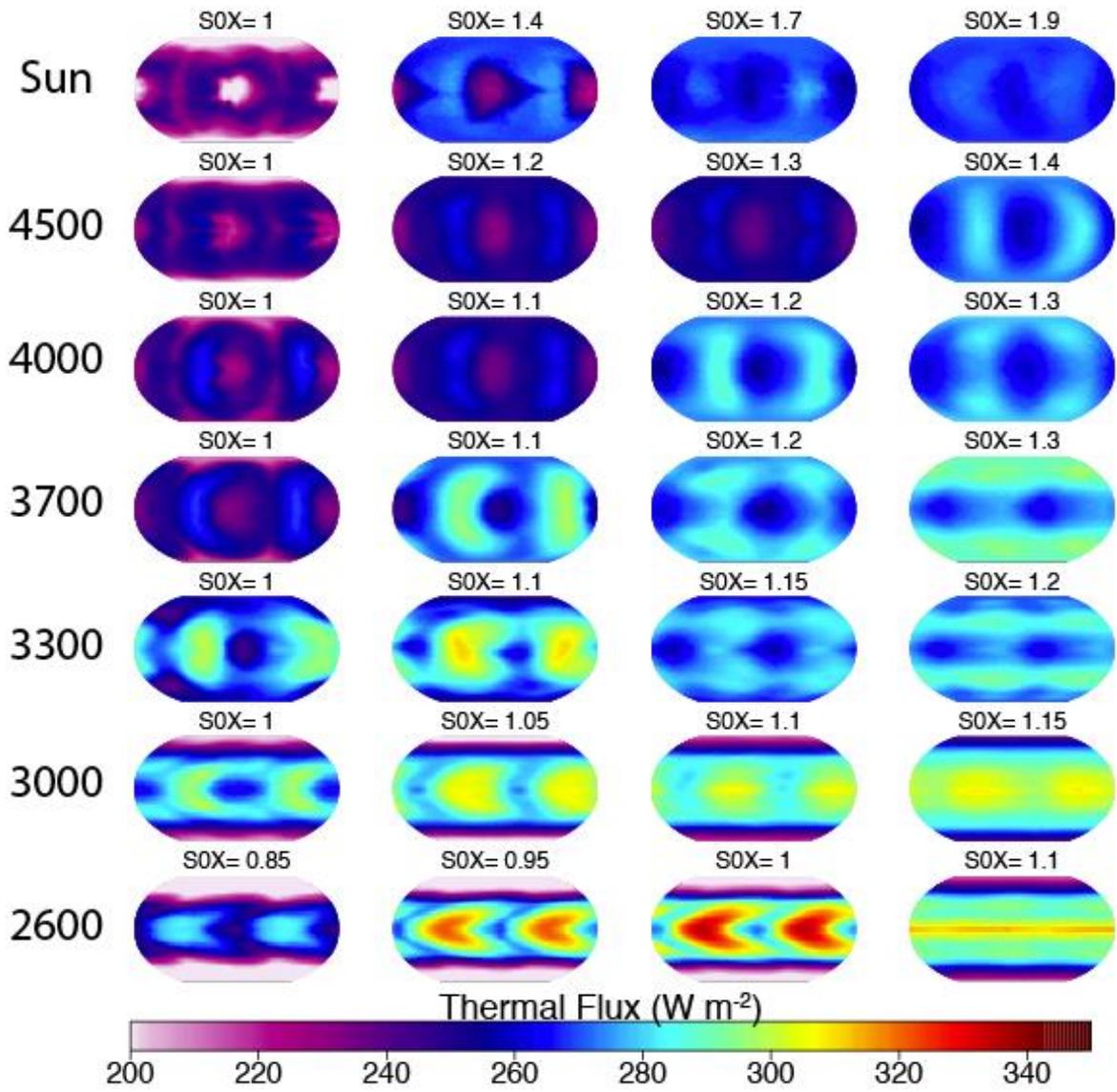

Figure S9. As in Figure S8, except results shown for 3:2 resonance planets.

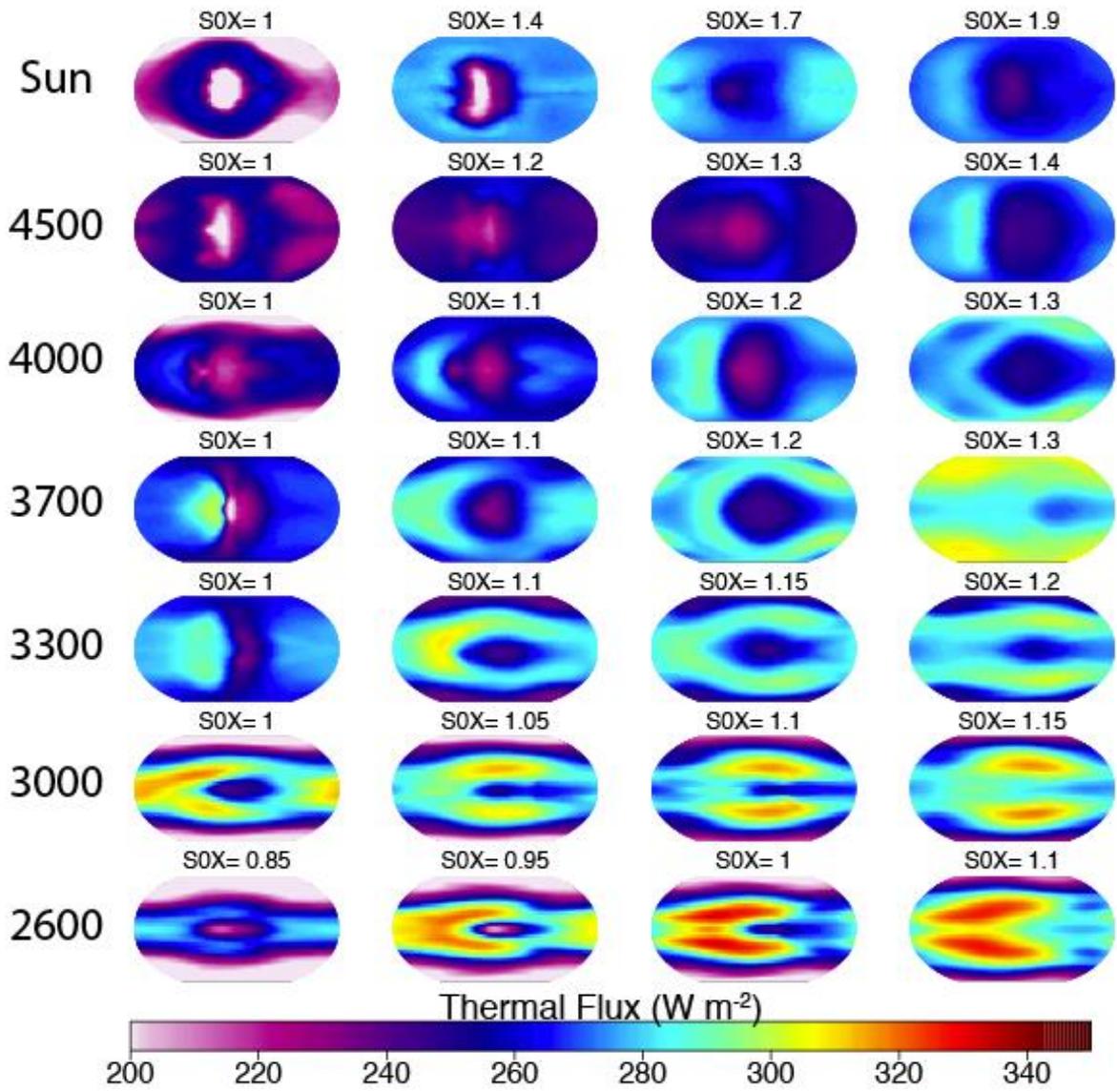

Figure S10. Figure S8-S9, except results shown for 2:1 resonance planets.

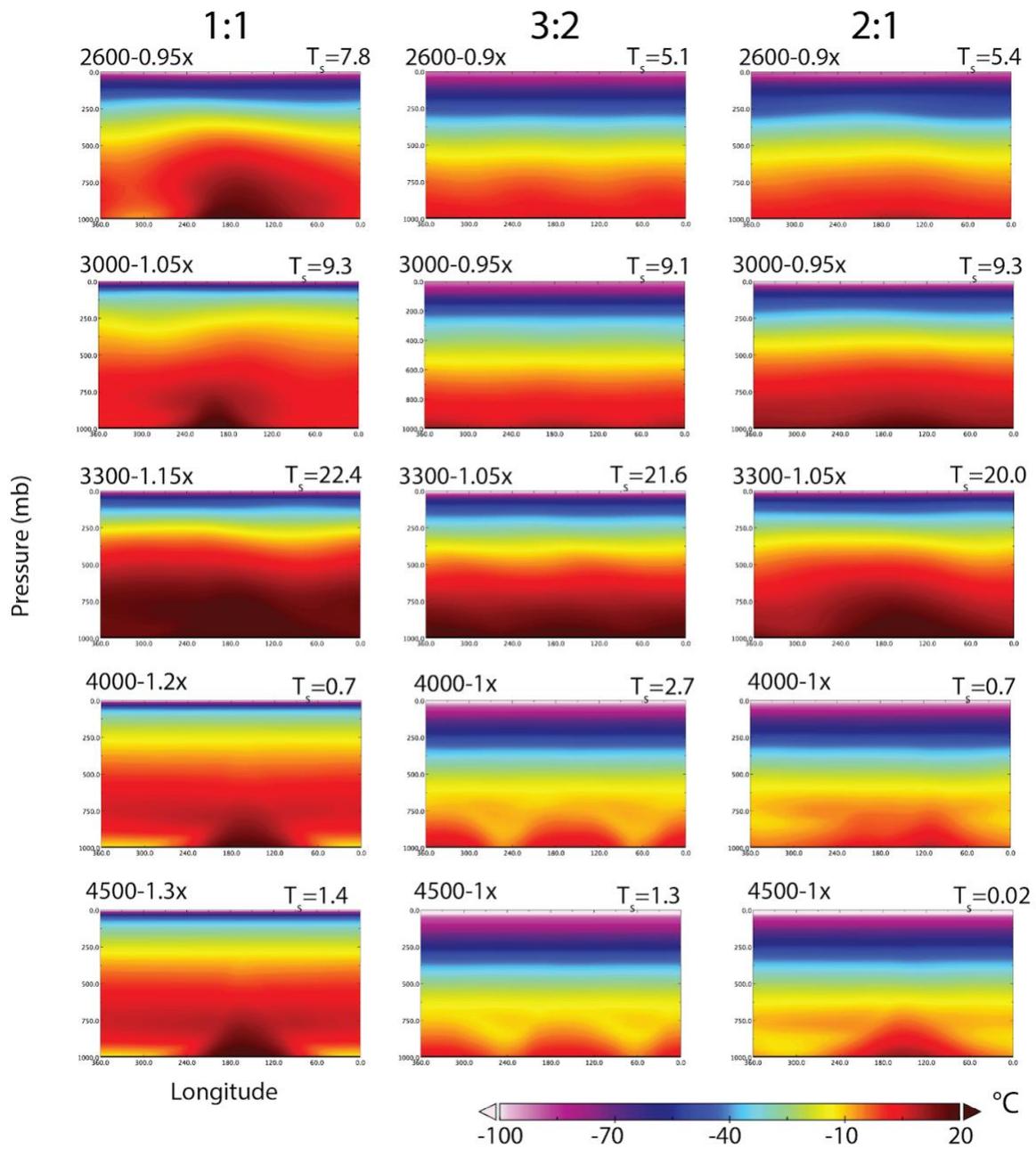

Figure S11. Vertical profiles of temperature (°C) vs. longitude for meridionally averaged (area-weighted) temperature between 30°N and 30°S. Panels are for select simulations (stellar type and flux labeled at top-left of each subplot) with common global-mean surface temperatures (labeled at top-right) for the (left column) 1:1, (middle column) 3:2, and (right column) 2:1 resonances. Stellar type is the same across each row.

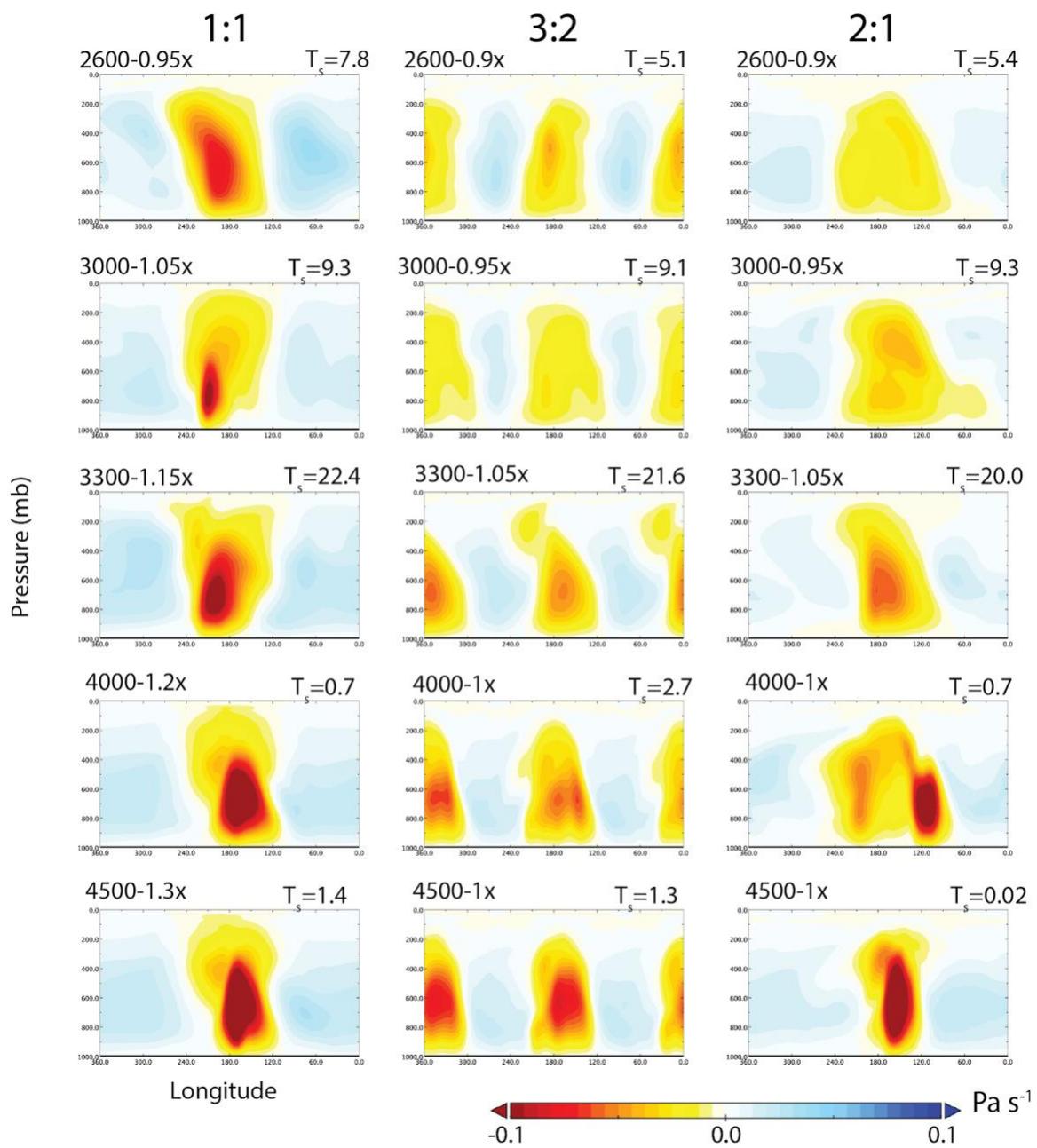

Figure S12. As in figure S11, for the same selected simulations, except results shown for vertical velocity, ω (Pa s⁻¹). Negative values (red) indicate upward motion and vice versa.